\documentclass[a4paper,12pt]{article}
\usepackage{latexsym}
\usepackage{amsmath,amssymb,amsxtra,amscd,amsthm}
\usepackage{bbm}
\usepackage{rotating}
\usepackage{color}
\usepackage{hyperref}
\numberwithin{equation}{section}
\allowdisplaybreaks
\newcommand{\p}{\partial}

\begin{document}

\setlength{\parindent}{0cm}
\setlength{\baselineskip}{1.5em}
\title{\bf Lectures on Mirror Symmetry \\and Topological String Theory}
\author{Murad Alim
\\
\small Jefferson Physical Laboratory, Harvard University, \\ \small Cambridge, MA 02138, USA }

\date{}
\maketitle

\abstract{These are notes of a series of lectures on mirror symmetry and topological string theory given at the Mathematical Sciences Center at Tsinghua University. The $\mathcal{N}=2$ superconformal algebra, its deformations and its chiral ring are reviewed. A topological field theory can be constructed whose observables are only the elements of the chiral ring. When coupled to gravity, this leads to topological string theory. The identification of the topological string A- and B-models by mirror symmetry leads to surprising connections in mathematics and provides tools for exact computations as well as new insights in physics. A recursive construction of the higher genus amplitudes of topological string theory expressed as polynomials is reviewed. }

\clearpage


\tableofcontents

\section{Introduction}
The study of supersymmetric gauge theories and string theory has been a very fruitful arena of interaction between mathematics and physics which has pushed the boundaries of knowledge and understanding on both sides. A key lesson learned on the physics side is that it is very useful to think of physical theories in terms of families, where members of a family of theories are related by deformations. This amounts to studying theories together with their space of couplings. Physical theories are usually formulated using a Lagrangian for the light degrees of freedom and their interactions. When the interactions are governed by a small coupling, perturbative techniques can be used to evaluate the correlation functions of the physical observables. Perturbation theory does however not take into account non-perturbative effects which are nontrivial field configurations with finite action which have to be considered as various backgrounds for perturbation theory and are in general hard to study. As the coupling of the theory is varied, perturbation theory breaks down as the non-perturbative effects become more and more important signaling that the original choice of light degrees of freedom is no longer valid. Hence, in different regions in the coupling space, perturbative and non-perturbative degrees of freedom may interchange. 

A great leap forward in the understanding of the interplay of perturbative and non-perturbative degrees of freedom and the physical manifestation of their interchange in terms of dualities followed the pioneering work of Seiberg and Witten (SW) \cite{Seiberg:1994rs}. In this work the exact low energy effective theory was determined including all non-perturbative effects using holomorphicity and the expected behavior at singularities in the space of the effective couplings due to physical particles becoming massless. The key insight was that the problem of determining the exact couplings could be mapped to an equivalent mathematical problem of determining periods of a curve. This auxiliary geometry could be given a physical meaning, identifying it with part of the Calabi-Yau (CY) threefold geometry of a type IIB string compactification \cite{Klemm:1996bj}, embedding this understanding into the larger framework of string dualities (See \cite{Lerche:1996xu} for a review). 

One particular physical duality which had emerged earlier is mirror symmetry. It was observed that an exchange of signs of some generators of the $\mathcal{N}=(2,2)$ superconformal algebra underlying some string compactifications leads to an isomorphism \cite{Dixon:1987ld}. This isomorphism identifies furthermore a ring of operators in the superconformal algebra, the chiral ring \cite{Lerche:1989uy}. The deformation problem and the isomorphism defined abstractly using the superconformal algebra can be given a mathematical meaning when this algebra is realized by nonlinear sigma models defined on different target spaces $Z$ and $Z^*$, see \cite{Witten:1991zz} and references therein.\footnote{By now there is a vast literature on the subject of mirror symmetry, see in particular \cite{MS1,Cox:1999ms,Hori:2003ms,Aspinwall:2009ms} and references therein for a more complete account of the developments leading to mirror symmetry.} Truncating the states of a representation of the superconformal algebra to those created by operators in the chiral ring and summing over all $2d$ topologies defines topological string theory, the two isomorphic versions are called the A- and the B-model. These probe K\"ahler and complex structure deformations of mirror Calabi-Yau threefolds $Z$ and $Z^*$.  The B-model is more accessible to computations since its deformations are the complex structure deformations of $Z^*$ which are captured by the variation of Hodge structure. Mirror symmetry is established by providing the mirror maps which are a distinguished set of local coordinates in a given patch of the deformation space.  These provide the map to the A-model, since they are naturally associated with the chiral ring. 

At special loci in the moduli space, the A-model data provides enumerative information of the CY $Z$. Using methods \cite{Greene:1990ud} to construct the mirror manifold of the quintic, it was thus possible \cite{Candelas:1990rm} to compute the data associated to the variation of Hodge structure in the B-model and make predictions of the Gromov-Witten (GW) invariants. The higher genus GW invariants can be resummed to give integer multiplicities of BPS states in a five-dimensional theory obtained from an M-theory compactification on $Z$ \cite{Gopakumar:1998ii,Gopakumar:1998jq}. Moreover, the special geometry governing the deformation spaces allows one to compute the prepotential $F_0(t)$ which governs the exact effective action of the four dimensional theories obtained from compactifying type IIA(IIB) string theory on $Z(Z^*)$, respectively. Various $4d$ gauge theories can be geometrically engineered \cite{Katz:1996fh,Katz:1997eq} and mirror symmetry can be used to compute their exact effective actions. (See also \cite{Mayr:1998tx} and references therein.)

The prepotential is the genus zero free energy of topological string theory, which is defined perturbatively in a coupling constant governing the higher genus expansion. The partition function of topological string theory indicating its dependence on local coordinates in the deformation space has the form:
\begin{equation}
\mathcal{Z}(t,\bar{t})=\exp \left(  \sum_{g} \lambda^{2g-2} \mathcal{F}^{(g)}(t,\bar{t}) \right) \, .
\end{equation} 
In Refs.~\cite{Bershadsky:1993ta,Bershadsky:1993cx}, Bershadsky, Cecotti, Ooguri and Vafa (BCOV) developed the theory and properties of the higher genus topological string free energies putting forward recursive equations, the holomorphic anomaly equations along with a method to solve these in terms of Feynman diagrams. For the full partition function these equations take the form of a heat equation \cite{Bershadsky:1993cx,Witten:1993ed} and can be interpreted \cite{Witten:1993ed} as describing the background independence of the partition function when the latter is interpreted as a wave function associated to the geometric quantization of $H^3(Z^*)$.

Using the special geometry of the deformation space a polynomial structure of the higher genus amplitudes in a finite number of generators  was proven for the quintic and related one parameter deformation families \cite{Yamaguchi:2004bt} and generalized to arbitrary target CY manifolds \cite{Alim:2007qj}, see also Refs.~\cite{Aganagic:2006wq,Grimm:2007tm,Hosono:2008ve}. The polynomial structure supplemented by appropriate boundary conditions enhances the computability of higher genus amplitudes. Moreover the polynomial generators are expected to bridge the gap towards constructing the appropriate modular forms for a given target space duality group which is reflected by the special geometry of the CY manifold.

The aim of these notes is to provide an accessible concise introduction into some of the ideas outlined above. The presentation is far from being self-contained and the topics which are covered represent a small sample of a vast amount of developments in this field, reference will be given to the original literature and to many of the excellent reviews on the subject. By the nature of it, the material presented here is very close to own contributions and in fact it relies heavily on Refs.~\cite{Alim:2007qj,Alim:2008kp,Alim:2009zz,Alim:2012ss}. 

The outline of these notes is as follows. In Sec.~\ref{sca} the $\mathcal{N}=2$ superconformal algebra and its chiral ring as well as its deformations are reviewed. It is discussed how to construct a bundle of states out of the subring of the chiral ring spanned by the deformation operators. Furthermore, the geometric realization of the $\mathcal{N}=(2,2)$ superconformal algebra as a nonlinear sigma model is described.
In Sec.~\ref{topstring} it is outlined how topological string theory can be obtained by first restricting the physical observables to the chiral ring and then coupling to gravity. The two inequivalent ways of doing so, the A- and B-model are discussed as well as the geometric interpretation of the chiral ring on both sides. Special geometry and its B-model realization in the variation of Hodge structure are reviewed. The quintic and its mirror are discussed as an example of constructing mirror geometries and finding the mirror map.
In Sec.~\ref{highergenus}, the BCOV anomaly equation and its recursive solution are reviewed. Using special geometry, a polynomial structure of the higher genus amplitudes can be proven and used for easier computations when the physically expected boundary conditions are implemented. An example of applying this structure to the quintic is given.

\section{The superconformal algebra and the chiral ring} \label{sca}
Mirror symmetry originates from representations of the $\mathcal{N}=(2,2)$ superconformal algebra, which refers to two copies, left- and right moving of the $\mathcal{N}=2$ superconformal algebra which is discussed in the following. The topological string A- and B-models which are identified by mirror symmetry refer to a truncation of the states of a representation of $\mathcal{N}=(2,2)$ superconformal algebra to the chiral ring \cite{Lerche:1989uy}. The following is based on Refs.~\cite{Alim:2009zz,Greene:1996cy,Witten:1991zz,Warner:1993zh,Aspinwall:2004jr,Aspinwall:2009ms,Fre:1995bc}.

\subsection{$\mathcal{N}=2$ superconformal algebra}
The $\mathcal{N}=2$ superconformal algebra is generated by the energy momentum tensor $T(z)$, which has conformal weight $h=2$, by two anti-commuting currents $G^{\pm}(z)$ of conformal weight $3/2$ and a $U(1)$ current $J(z)$ under which the $G^{\pm}(z)$ carry charge $\pm$ and $z$ is a local coordinate on the $2d$ worldsheet. These currents satisfy the following operator product expansions (OPE):
\begin{eqnarray}\label{scaope}
G^{\pm}(z)G^{\mp}(w)&=&\frac{\frac{2}{3}c}{(z-w)^3} \pm \frac{2J(w)}{(z-w)^2}+ \frac{2T(w)\pm \partial_wJ(w)}{(z-w)} + \dots \,,\\
J(z) \,G^{\pm}(w)&=& \pm \frac{G^{\pm}(w)}{(z-w)}+\dots\,,\\
J(z)\,J(w)&=&\frac{\frac{1}{3}c}{(z-w)^2}+\dots\,,\\
T(z) \,J(w) &=& \frac{J(w)}{(z-w)^2}+\frac{\partial_w J(w)}{(z-w)}+\dots\,,\\
T(z)\, G^{\pm}(w)&=& \frac{\frac{3}{2}G^{\pm}(w)}{(z-w)^2}+\frac{\partial_w G^{\pm}(w)}{(z-w)}+\dots\,,\\
T(z)\,T(w)&=& \frac{\frac{1}{2}c}{(z-w)^4}+\frac{2T(w)}{(z-w)^2}+\frac{\partial_wT(w)}{(z-w)}+\dots\,,
\end{eqnarray}
where the dots refer to the addition of regular terms in the limit $z\rightarrow w$, $c$ is the central charge. The  boundary conditions which must be imposed for the currents $G^{\pm}(z)$ can be summarized as follows
\begin{equation}
G^{\pm}(e^{2\pi i}z)=- e^{\mp 2\pi i a} G^{\pm}(z)  \, ,
\end{equation}
with a continuous parameter $a$ which lies in the range $0 \le a < 1$. For integral and half integral $a$ one recovers anti-periodic and periodic boundary conditions corresponding to the Ramond and Neveu-Schwarz sectors, respectively. The currents can be expanded in Fourier modes
\begin{equation}\label{modes}
T(z)= \sum_{n} \frac{L_n}{z^{n+2}}\, ,\quad G^{\pm}(z)= \sum_{n} \frac{G^{\pm}_{n \pm a}}{z^{n\pm a+\frac{3}{2}}}\, ,\quad J(z)=\sum_n \frac{J_n}{z^{n+1}} \, .
\end{equation}
The $\mathcal{N}=2$ superconformal algebra can be expressed in terms of the operator product expansion of the currents or by the commutation relations of their modes. The latter reads:
\begin{eqnarray}
\left[ L_m, L_n \right] &=& (m-n) L_{m+n} +\frac{c}{12} m (m^2-1) \delta_{m+n,0} \, , \nonumber \\
\left[  J_m,J_n\right] &=& \frac{c}{3}  m \delta_{m+n,0}\, ,\nonumber\\
\left[ L_n, J_m\right] &=& -m J_{m+n} \, ,\nonumber\\
\left[L_n,G_{m\pm a}^{\pm} \right] &=& \left(\frac{n}{2}-(m\pm a)\right) G^{\pm}_{m+n\pm a}\, , \nonumber\\
\left[ J_n,G_{m\pm a}^{\pm}\right] &=& \pm G_{n+m\pm a}^{\pm} \, ,\nonumber\\
\left\{  G_{n+a}^+,G^-_{m-a}\right\}&=& 2 L_{m+n} + (n-m+2 a)J_{n+m}+\frac{c}{3} \left((n+a)^2-\frac{1}{4}\right)\delta_{m+n,0}\, .
\end{eqnarray}
The algebras obtained for every value of the continuous parameter $a$  are isomorphic. This isomorphism induces an operation on the states which is called \emph{spectral flow}. This operation shows the equivalence of the Ramond (R) and Neveu-Schwarz (NS) sectors as the states in each sector are continuously related by the flow. Moreover, as NS and R sectors give rise to space-time bosons and space-time fermions respectively, this isomorphism of the algebra induces space-time supersymmetry.

\subsection{Chiral ring}
The representation theory of the $\mathcal{N}=2$ superconformal algebra is equipped with an interesting additional structure, namely a finite sub-sector of the operators creating the highest weight states carries an additional ring structure which is discussed in the following.

 The unitary irreducible representations of the algebra can be built up from highest weight states by acting on these with creation operators which are identified with the negative modes in (\ref{modes}). Similarly all the modes with positive indices can be thought of as annihilation operators which lower the $L_0$ eigenvalue of a state. A highest weight state is thus one which satisfies,
\begin{equation}
 L_n |\phi \rangle =0\, , \quad G^{\pm}_r |\phi\rangle=0\, ,\quad J_m|\phi\rangle =0 \,,\quad n,r,m>0\,.
\end{equation}
The zero index modes $L_0$ and $J_0$ modes can be used to label the states by their eigenvalues
\begin{equation}
 L_0|\phi\rangle =h_{\phi}|\phi \rangle \,, \quad J_0 |\phi \rangle = q_{\phi} |\phi\rangle\,.
\end{equation}
In the Ramond sector there are furthermore the modes $G_0^{\pm}$. If a state is annihilated by these then it is called a  Ramond ground state. A highest weight state is created by a primary field $\phi$
\begin{equation}
 \phi |0\rangle = |\phi\rangle\,.
\end{equation}
The subset of primary fields which will be of interest is constituted of the \emph{chiral primary} fields. States which are created by those satisfy furthermore
\begin{equation}\label{chiral}
 G^{+}_{-1/2} |\phi\rangle =0\,.
\end{equation}
The name anti-chiral primary will be used for the primary fields annihilated by $G^-_{-1/2}$. In combination with the representations of the anti-holomorphic currents $\overline{G}^{\pm}$ this leads to the notions of $(c,c),(a,c),(a,a)$ and $(c,a)$ primary fields, where $c$ and $a$ stand for chiral and anti-chiral and the pair denotes the conditions in the holomorphic and anti-holomorphic sectors. Considering
\begin{equation}
 \langle \phi| \{G^-_{1/2},G^+_{-1/2}\} |\phi\rangle = || G^+_{-1/2} |\phi\rangle||^2= \langle \phi|2L_0-J_0|\phi\rangle\,\ge 0\,,
\end{equation}
implies 
\begin{equation}
h_{\phi}\ge \frac{q_{\phi}}{2} \, , 
\end{equation}
with equality holding for chiral states. This property of chiral primary states is an analog of the BPS bound for physical states. Now looking at the operator product expansion of two chiral primary fields 
$\phi$ and $\chi$
\begin{equation}
 \phi(z) \chi(w) =\sum_i (z-w)^{h_{\psi_i} -h_{\phi}-h_{\chi}} \psi_i\, ,
\end{equation}
the $U(1)$ charges add $q_{\psi_i}=q_{\phi}+q_{\chi}$ and hence $h_{\psi_i}\geq h_{\phi}+h_{\chi}$. The operator product expansion has thus no singular terms and the only terms which survive in the expansion when $z\rightarrow w$ are the ones for which $\psi_i$ is itself chiral primary. It is thus shown that the chiral primary fields give a closed non-singular ring under operator product expansion. Furthermore, one can show the finiteness of this ring by considering
\begin{equation}
 \langle \phi| \{G^-_{3/2},G^+_{-3/2}\} |\phi\rangle = \langle \phi|2L_0-3 J_0+\frac{2}{3}c|\phi\rangle \geq 0\,,
\end{equation}
to see that the conformal weight of a chiral primary is bounded by $c/6$. 

For the worldsheet description of certain string compactifications two copies of the $\mathcal{N}=2$ superconformal algebra are required, these are usually referred to as left and right moving or holomorphic and anti-holomorphic, the currents of the latter will be denoted by $\overline{T},\overline{G}^{\pm},\overline{J}$. The algebra is then referred to as the $\mathcal{N}=(2,2)$ superconformal algebra. 

In the $\mathcal{N}=(2,2)$ superconformal algebra there are now four finite rings, depending on the different combinations of chiral and antichiral in the right and left moving algebras, this gives the $(c,c),(a,c),(a,a)$ and $(c,a)$ rings, one sees that the latter two are charge conjugates of the first two. The relation between charge and conformal weight for an anti-chiral primary becomes $h_{\psi}=-\frac{q_{\psi}}{2}.$ Denoting the set of chiral primary fields by $\phi_i$ where the index $i$ runs over all chiral primaries, the ring structure can be formulated as follows\footnote{Formulas for products of operators are understood to hold within correlation functions.}
\begin{equation}\label{chiralring}
 \phi_i \phi_j = C_{ij}^k \phi_k \,.
\end{equation}

\subsection{Deformation families}
Mirror symmetry is a symmetry relating the deformation family of the $(a,c)$ chiral ring with the deformation family of the $(c,c)$ chiral ring. To make this statement more precise the deformation family of the $(c,c)$ chiral ring will be discussed in the following. Deformations of a conformal field theory are achieved by adding marginal operators to the original action, these are operators having conformal weight $h+\overline{h}=2$. In the following spinless operators will be studied which have $h=1$, $\overline{h}=1$. The operators which maintain their $h=\overline{h}=1$ conformal weights after perturbation of the theory are called truly marginal operators. Such operators can be constructed from the chiral primary operators in two steps. For instance in the $(c,c)$ ring, starting from an operator of charge $q=\overline{q}=1$, $h=\overline{h}=1/2$ one can first construct
\begin{equation}
 \phi^{(1)}(w,\overline{w})=\left[  G^-(z),\phi(w,\overline{w})\right]=\oint dz\, G^{-}(z)\phi(w,\overline{w})\, ,
\end{equation}
 which now has $h=1,q=0$. In the next step
\begin{equation}
\phi^{(2)}(w,\overline{w})=\left\{ \overline{G}^{-}(z),\phi^{(1)}(w,\overline{w})\right\}=\oint d\overline{z}\, \overline{G}^{-}(\overline{z}) \phi^{(1)}(w,\overline{w})\, ,
\end{equation}
which has $h=\overline{h}=1$ and zero charge and is hence a truly marginal operator and can be used to perturb the action of the theory
\begin{equation}
\delta S = t^i \int \phi^{(2)}_i + \overline{t}^{\bar{\imath}} \int \phi^{(2)}_{\bar{\imath}}\, , \quad i=1,\dots,n\, ,
\end{equation}
where a priori also the deformations coming from the $(a,a)$ operators are included and $n=\textrm{dim} \mathcal{H}^{(1,1)}$ denotes the dimension of the subspace of the Hilbert space of the theory spanned by the states which are created by the charge $(1,1)$ operators. A similar construction can be done for the $(a,c)$ chiral ring. The superscript notation is borrowed from topological field theories where an analogous construction gives the two form descendants which can be used to perturb the topological theory, see for example \cite{Dijkgraaf:1990qw}. The deformations constructed in this fashion span a deformation space $\mathcal{M}$, the moduli space of the SCFT.

\subsection{The deformation subring}
For the study of the geometric realization of the equivalence of the SCFTs a holomorphic vector bundle $V_{\mathbbm{C}}$ over the deformation space $\mathcal{M}$ is introduced. Its fiber corresponds to a subset of the ground-states of the theory. Its importance lies in the fact that the ground-states of the theory do not change over the space of deformations \cite{Hori:2003ms}. However, there is a certain way of splitting the bundle which varies smoothly over the moduli space. This bundle collects the states of the theory which are created by the sub-ring of the chiral ring spanned by the charge $(1,1)$ operators.
A basis for this sub-ring is denoted by $(\phi_0,\phi_i,(\phi_D)_i,(\phi_D)_0), \, i=1,\dots,n$. $\phi_0$ is the identity operator of charge $(0,0)$ and $(\phi_D)_i$ are the charge $(2,2)$ fields  and $(\phi_D)_0$ is an element of charge $(3,3)$.

The threepoint function on the sphere is the correlator of 3 chiral fields of charge $(1,1)$:\footnote{The charges of all operators in a correlation function in the topological theory have to add up to $(3,3)$ due to an anomaly in the $U(1)$ current (See for example \cite{Warner:1993zh,Aspinwall:2009ms}). }
\begin{equation}
C_{ijk}= \langle \phi_i \phi_j \phi_k \rangle\, ,\quad i,j,k=1,\dots \textrm{dim}\mathcal{H}^{1,1}\, .
\end{equation}
Using the chiral ring relation (\ref{chiralring}) this allows the following manipulation:
\begin{equation}
C_{ijk}=\langle \phi_i \phi_j \phi_k \rangle = \langle \phi_i C_{jk}^m  (\phi_D)_m \rangle=C_{jk}^m  \langle \phi_i (\phi_D)_m \rangle= C_{jk}^m \, \eta_{im}\, ,
\end{equation}
defining the topological metric $\eta_{im}$ which gives a pairing between the charge $(1,1)$ and charge $(2,2)$ operators. This metric can be used to raise and lower indices and identify the structure constants of the chiral ring with the threepoint function.\footnote{See Ref.~\cite{Dijkgraaf:1990qw} for a review with more details on the topological metric.}

Using $\phi^i=\eta^{ij}(\phi_D)_j$ and $\langle (\phi_D)_0\rangle=1\,,\eta_{00}=1$, the ring structure can now be put in matrix form 
\begin{equation} \label{chiralringmat}
\phi_i  \left( \begin{array}
         {c}\phi_0\\\phi_j\\ \phi^j\\ \phi^0
        \end{array} \right)
=
\left(\begin{array}{cccc}
       0 &\delta_i^k&0&0 \\ 0&0&C_{ijk}&0 \\ 0&0&0&\delta_i^j \\0&0&0&0 
      \end{array}
\right) \left( \begin{array}
         {c}\phi_0\\\phi_k\\ \phi^k\\ \phi^0
        \end{array} \right) \,.
\end{equation}
The states created by this sub-ring are organized in a vector bundle
\begin{equation}
 \mathcal{V}=\mathcal{H}^{0,0} \oplus \mathcal{H}^{1,1} \oplus \mathcal{H}^{2,2}\oplus \mathcal{H}^{3,3}\, ,
\end{equation}
where $\mathcal{H}^{(i,i)}$ denotes the subspace of the Hilbert space of states created by charge $(i,i)$ operators. The splitting of the bundle is given by the charge grading. The variation of this grading over the moduli space $\mathcal{M}$ and its geometric realization is going to be a central theme in the study of mirror symmetry.
The operator-state correspondence can be used to obtain a basis of the vector bundle from the basis of the chiral ring operators. Denoting by $|e_0\rangle \in \mathcal{H}^{0,0}$ the unique ground state of charge $(0,0)$ up to scale, a basis for $\mathcal{V}$ can be obtained as follows
\begin{equation}
 |e_i\rangle = \phi_i |e_0\rangle \, , \quad |e^i\rangle=\phi^i|e_0\rangle\, ,\quad |e^0\rangle=\phi^0 |e_0\rangle \, . 
\end{equation}
The metric on this bundle is given by the topological metric
\begin{equation}
 \langle e_a|e^b \rangle = \delta_a^b\, \quad a,b=0,\dots,n\,.
\end{equation}
The representation of the chiral ring in this basis reads
\begin{equation} \label{topGM}
\phi_i  \left( \begin{array}{c}
       |e_0 \rangle\\|e_j \rangle\\|e^j \rangle\\|e^0 \rangle
       \end{array} \right)
=\underbrace{\left(\begin{array}{cccc}
      0 &\delta_i^k&0&0 \\ 0&0&C_{ijk}&0 \\ 0&0&0&\delta_i^j \\0&0&0&0 
     \end{array}
\right)}_{:=C_i} \left( \begin{array}{c}
       |e_0 \rangle\\|e_k \rangle\\|e^k \rangle\\|e^0 \rangle
       \end{array} \right)\,.
\end{equation}
This structure will turn out to be crucial for the understanding of mirror symmetry. In terms of coordinates on the moduli space $\mathcal{M}$ an insertion of the chiral field $\phi_i$ is equivalent to an infinitesimal displacement in moduli space and hence can be obtained by a derivative $\frac{\partial}{\partial t^i}$. Denoting the matrix on the right hand side by $C_i$, the whole equation can be read as a connection on the bundle $\nabla=\partial_i - C_i$ which is flat.\footnote{See \cite{Bershadsky:1993cx,Hori:2003ms,Walcher:2007tp} for more details.} 
\begin{equation}
\left[ \nabla_i,\nabla_j \right]=0\,.
\end{equation}
This connection is called the Gauss-Manin connection since the geometric realization of it on the B-model side coincides with the Gauss-Manin connection defined in the variation of Hodge structure.

\subsection{Geometric realization of SCFT}\label{geometricrealization}
A geometric realization of the $\mathcal{N}=(2,2)$ SCFT is provided by the nonlinear sigma model \cite{Witten:1988xj,Witten:1991zz}, see also the reviews in Refs.~\cite{Hori:2003ms,Aspinwall:2004jr,Aspinwall:2009ms}. The focus here is on the geometric realization of the $(a,c)$ and $(c,c)$ rings. The nonlinear sigma model that will be considered is a field theory of bosons $\phi$ and fermions $\psi,\chi$ living on a $2d$ Riemann surface $\Sigma$, being related by supersymmetry where the bosonic fields are considered as coordinates of some target space $X$, i.e., $\phi:\Sigma\rightarrow X$. In order for the theory to have $\mathcal{N}=(2,2)$ supersymmetry the target space spanned by the bosonic fields has to be a K\"ahler manifold, which in turn allows one to split its tangent bundle $TX=T^{1,0}X\oplus T^{0,1}X$. The action of the theory is (as given in Refs.~\cite{Witten:1991zz,Aspinwall:2009ms})\footnote{The notation is chosen differently from Refs.~\cite{Witten:1991zz,Aspinwall:2009ms}, in order to be more suitable for the discussion of higher genus topological strings, where the notation of Ref.~\cite{Bershadsky:1993cx} is adopted. The superscript $\pm$ refers to the $U(1)$ charge of the superconformal algebra currents whereas the bar on top of the expressions refers to the right-moving sector. For the fermions, the left moving fermions are denoted by $\psi$ whereas the right moving fermions are denoted by $\chi$.} 
\begin{eqnarray}
 S&=& \frac{1}{4\pi}\int_{\Sigma} d^2z (  g_{i\bar{\jmath}} \left( \partial_{z} \phi^i \partial_{\bar{z}}\phi^{\bar{\jmath}} +\partial_{\bar{z}} \phi^i \partial_{z}\phi^{\bar{\jmath}}  \right)+ B_{i\bar{\jmath}} \left( \partial_{z} \phi^i \partial_{\bar{z}}\phi^{\bar{\jmath}} -\partial_{\bar{z}} \phi^i \partial_{z}\phi^{\bar{\jmath}}  \right)  \nonumber\\&+&i g_{i\bar{\jmath}} \psi^i D_z\psi^{\bar{\jmath}} +i g_{i\bar{\jmath}} \chi^i D_{\bar{z}} \chi^{\bar{\jmath}}+ R_{i\bar{k}j\bar{l}} \psi^i \psi^{\bar{k}} \chi^j \chi^{\bar{l}}  ) \, .
\end{eqnarray}
Denoting by $K$ the canonical bundle on $\Sigma$ the fermions are sections of
\begin{eqnarray}
\psi^i \in \Gamma(K^{1/2}\otimes \phi^* T^{1,0}X)\,, \quad  && \psi^{\bar{\imath}} \in \Gamma(K^{1/2}\otimes \phi^* T^{0,1} X)\,,\nonumber\\
\chi^i \in \Gamma(\overline{K}^{1/2}\otimes \phi^* T^{1,0}X)\, , \quad&&   \chi^{\bar{\imath}} \in \Gamma(\overline{K}^{1/2}\otimes \phi^* T^{0,1}X)\, .
\end{eqnarray}
The covariant derivative in the action is with respect to these bundles. The fermions $\psi$ and $\chi$ correspond to the left and right moving sectors. The action is conformally invariant if the Ricci tensor is vanishing, together with the K\"ahler condition this yields a Calabi-Yau target space.
The supersymmetries are given by the following transformations of the fields:
\begin{eqnarray} \label{susytrafo1}
\delta \, \phi^i &=& i \varepsilon^- \, \psi^i + i \bar{\varepsilon}^- \, \chi^i\,,  \\ \label{susytrafo2}
\delta \, \phi^{\bar{\imath}} &=& i \varepsilon^+ \, \psi^{\bar{\imath}} + i \bar{\varepsilon}^+ \, \chi^{\bar{\imath}}\,,  \\ \label{susytrafo3}
\delta \psi^i &=& -\varepsilon^+ \partial \phi^i- i \bar{\varepsilon}^- \chi^j\,\Gamma_{jk}^i\, \psi^k\, ,\\  \label{susytrafo4}
\delta \psi^{\bar{\imath}} &=& -\varepsilon^- \partial \phi^i- i \bar{\varepsilon}^+ \chi^{\bar{\jmath}}\,\Gamma_{\bar{\jmath}\bar{k}}^{\bar{\imath}} \,\psi^{\bar{k}}\, ,\\ \label{susytrafo5}
\delta \chi^i &=& -\bar{\varepsilon}^+ \bar{\partial} \phi^i- i \varepsilon^- \psi^j\,\Gamma_{jk}^i \,\chi^k\, ,\\  \label{susytrafo6}
\delta \chi^{\bar{\imath}} &=& -\bar{\varepsilon}^- \bar{\partial} \phi^{\bar{\imath}}- i \varepsilon^+ \psi^{\bar{\jmath}}\,\Gamma_{\bar{\jmath}\bar{k}}^{\bar{\imath}} \,\chi^{\bar{k}}\, ,
\end{eqnarray}
where $\varepsilon^-,\bar{\varepsilon}^-,\varepsilon^+,\bar{\varepsilon}^+$ are sections of $K^{-1/2},\overline{K}^{-1/2},K^{-1/2},\overline{K}^{-1/2}$, respectively.
The geometric realization of the currents (\ref{scaope}) is given by:
\begin{eqnarray}
T&=& -g_{i\bar{\jmath}} \partial_z \phi^i\, \partial_z \phi^{\bar{\jmath}} +\frac{1}{2} g_{i\bar{\jmath}} \psi^i \partial_z \psi^{\bar{\jmath}}+\frac{1}{2} g_{i\bar{\jmath}} \psi^{\bar{\jmath}} \partial_z \psi^{i}\, ,\\
G^+&=&\frac{1}{2} g_{i\bar{\jmath}} \psi^i \partial_z \phi^{\bar{\jmath}}\, ,\\
G^-&=&\frac{1}{2} g_{i\bar{\jmath}} \psi^{\bar{\jmath}} \partial_z \phi^{i}\, ,\\
J&=&\frac{1}{4}g_{i\bar{\jmath}} \psi^{i} \psi^{\bar{\jmath}}  \, ,
\end{eqnarray}
and similarly for $\overline{T},\overline{G}^+,\overline{G}^-,\overline{J}$.
The supersymmetry variation in terms of the charges of the supersymmetry currents reads:\footnote{$Q^{\pm}=\frac{1}{2\pi i} \oint G^{\pm} dz\,,\quad \overline{Q}^{\pm}=\frac{1}{2\pi i}\oint \overline{G}^{\pm} d\overline{z}.$}
\begin{equation}
\delta= \varepsilon^+ \, Q^+ + \varepsilon^-\, Q^- + \overline{\varepsilon}^+ \, \overline{Q}^+ + \overline{\varepsilon}^- \, \overline{Q}^-\,.
\end{equation}


\section{Topological string theory and mirror symmetry} \label{topstring}
The geometric realization of the isomorphic $(a,c)$ and $(c,c)$ chiral rings of the superconformal algebra are studied using topological string theory which is a $2d$ topological field theory coupled to gravity.


\subsection{Topological field theory}
In the following some features of topological sigma models of cohomological type \cite{Witten:1988xj} will be outlined. These are sigma models, i.e.,  theories of maps $\phi:\Sigma_g \rightarrow X$ into some target space $X$, the case of interest is where $\Sigma_g$ denotes a Riemann surface of genus $g$. The maps are realized physically as bosons and fermions which are related by supersymmetry. The additional structure of cohomological topological field theories is the existence of a Grassmann, scalar symmetry operator $\mathcal{Q}$, which is sometimes called the BRST operator, with the following properties:
\begin{itemize}\label{topfield}
\item $\mathcal{Q}$ is nilpotent $\mathcal{Q}^2=0\,.$
\item The energy momentum tensor is $\mathcal{Q}$ exact,
$$ T_{\mu \nu}=\{ \mathcal{Q},G_{\mu \nu}\}\,.$$
\end{itemize}
A direct consequence of the last property is that the correlation functions do not depend on the two dimensional metric. To see this one notices that a variation with respect to the metric is equivalent to inserting the energy momentum tensor in the correlators, this one being exact and $\mathcal{Q}$ being a symmetry of the theory shows the metric independence. Furthermore, physical states of the theory correspond to cohomology classes of the operator $\mathcal{Q}$ and the semi-classical approximation for evaluating the path integral is exact. Further background as well as derivations of these properties can be found in \cite{Vonk:2005yv,Hori:2003ms}.

\subsection{Twisting and topological string theory}\label{deftopstring}
A topological field theory can be constructed from the $\mathcal{N}=2$ superconformal field theories whose states are those created by operators in the chiral ring. This is achieved by treating the annihilators of the states in question as BRST operators and considering only those states as physical states which are annihilated by the BRST operator. From the definition of a chiral state in Eq.~(\ref{chiral}) and its anti-chiral counterpart it is clear that if one wants to restrict to the $(a,c)/(c,c)$ chiral rings then the operators in question should be:
\begin{equation} 
\begin{array}{cc}
Q_A=G_{-1/2}^- + \overline{G}_{-1/2}^{+}, & Q_B=\quad G_{-1/2}^+ + \overline{G}^+_{-1/2}. \\ 
(a,c)&\quad (c,c)
\end{array}
\end{equation}
These operators square to zero in the algebra $(Q_{A/B})^2=0$ fulfilling the first requirement of (\ref{topfield}). The energy momentum tensor using either operator is however not $Q-$exact and furthermore the  operators are fermionic.

Topological string theory is obtained by coupling the resulting theory to $2d$ gravity which involves an integration over all $2d$ geometries and metrics. The two dimensional geometries are the Riemann surfaces which are organized by their genus $g$. One has in particular to be able to define the BRST operator globally on every such Riemann surface. The charges corresponding to the SCFT $G^{\pm}$ currents are however fermions. In order to have a global well defined charge on every Riemann surface the \emph{topological twist} is introduced. Since the spin of a current is determined by its conformal weight, which appears in the operator product expansion with the energy momentum tensor, the idea of the twist is to modify the energy momentum tensor. The new energy momentum tensor is a combination of the old one and the $U(1)$ current $J$. Depending on which ring one wants to restrict to, this is achieved by
\begin{equation} 
\begin{array}{c|c}\label{twist}
A&B\\
\hline
\\
T\rightarrow T+\frac{1}{2}\partial J& T\rightarrow T-\frac{1}{2}\partial J \\ 
\\
\overline{T}\rightarrow \overline{T}-\frac{1}{2}\overline{\partial} \overline{J}&\overline{T}\rightarrow \overline{T}-\frac{1}{2}\overline{\partial} \overline{J}\\
\\
\hline
(a,c)&(c,c)
\end{array}
\end{equation}
yielding Grassmann valued scalars $Q_A$ and $Q_B$.
Restricting to the chiral ring for example one now has:
\begin{equation}
T_{\textrm{top}}=T-\frac{1}{2}\partial \, J\,,
\end{equation} 
from the OPE (\ref{scaope}) it is possible to see that $G^+$ has now conformal weight 1. And furthermore one has:
\begin{equation}
T_{\textrm{top}}=\left\{ Q^+,G^-\right\}\,,
\end{equation}
provided that the $U(1)$ charge in correlators is $c/3$, where $c$ is the central charge. For the geometric realizations of the superconformal algebra as nonlinear sigma models into a CY threefold target space $c=9$.
The cohomology of these operators gives the desired restriction to the finite rings. As the notation suggests restricting to the $(a,c)$ ring gives the \emph{A-model} and restricting to the $(c,c)$ ring gives the \emph{B-model}. The coupling to gravity is achieved by defining a correlator for every genus of a Riemann surface and integrating the correlator over the moduli space of Riemann surfaces. The moduli space of a Riemann surface at genus $g$ is described by $3g-3$ Beltrami differentials $\mu^z_{\bar{z}}$, these are anti-holomorphic forms taking values in the holomorphic tangent bundle $\mu_a \in H^{0,1}(\Sigma_g,T \Sigma_g)$ and their complex conjugates $\overline{\mu}^{\overline{z}}_z$. These differentials are contracted with the two currents which get their conformal weight augmented to $h=2 \,(\bar{h}=2)$. This construction is motivated from the bosonic string where the anti-ghosts are contracted with the Beltrami differentials in a similar manner. The genus $g$ amplitudes or free energies are now defined by (here for the B-model and $g>1$)\footnote{The amplitudes at $g=0,1$ need separate treatment, in particular $\mathcal{F}^0$ denotes the prepotential and can be calculated from the geometry of the deformation bundle.}
\begin{equation}
 \mathcal{F}^g= \int_{\mathcal{M}_g} [dm \,d\bar{m}] \langle \prod_{a=1}^{3g-3}   (\int_{\Sigma} \mu_a G^-) (\int_{\Sigma} \mu_{\overline{a}} \overline{G}^-)\rangle_{\Sigma_g}\,,
\end{equation}
where $\langle \dots \rangle_{\Sigma_g}$ denotes the CFT correlator and $dm d\overline{m}$ are dual to the Beltrami differentials.\footnote{More details on topological strings can be found in \cite{Hori:2003ms,Marino:2004uf,Marino:2004eq,Neitzke:2004ni,Marino:2005sj,Vonk:2005yv,Pioline:2008zz}. }


\subsection{Mirror symmetry}
Mirror symmetry is the identification of the topological field (string-) theories which have as their physical states the $(a,c)$ and the $(c,c)$ chiral rings. Its origin in the superconformal algebra is the sign flip of the left-moving $U(1)$ current $J\leftrightarrow -J$. The power of mirror symmetry comes from the fact that the geometrical interpretation of the deformations and the chiral rings in the non-linear sigma model realizations are very different. 
The deformation of the A-model seeing the $(a,c)$ chiral ring are deformations of the complexified K\"ahler form of a CY threefold $Z$. The physical coordinate on the moduli space of theories is\footnote{It is assumed that $h^{2,0}(Z)=0\,.$}
\begin{equation}
t^a=\int_{C^a} B+i\mathcal{J}\, ,\quad a=1,\dots,h^{1,1}(Z)\,,
\end{equation}
where $C^a\in H_2(Z), B \in H^2(Z,\mathbbm{C})/H^{2}(Z,\mathbbm{Z})$ and $\mathcal{J}$ is the K\"ahler form.

The deformations on the B-side are complex structure deformations of a CY $Z^*$. The good physical coordinates are certain periods of the unique holomorphic $(3,0)$ form $\tilde{\Omega}$, with a certain normalization
\begin{equation}
t^a=\int_{\alpha^a} \tilde{\Omega}\,,\quad a=1,\dots,h^{2,1}(Z^*)\,,
\end{equation}
with $\alpha^a\in H_3(Z^*)$. The appropriate choice of coordinates on the moduli space of complex structures as well as the normalization of $\tilde{\Omega}$ is determined by the geometrical realization of the chiral ring and will be discussed later in these lectures. Mirror symmetry is about identifying mirror families of CY spaces $Z_t$ and $Z^*_{t^*}$ such that for every member of the family $h^{1,1}(Z)=h^{2,1}(Z^*)$ and the deformations match. The identification of the right physical coordinates on the moduli spaces of the two sides is achieved by realizing the geometrical counterpart of the deformation subring of the chiral ring on both sides.


\subsection{ A-model}
\subsubsection{\it A-twist}
The effect of the twist (\ref{twist}) on the non-linear sigma model with target space $Z$ is changing the bundles of which the fermions are sections (See \cite{Witten:1991zz,Hori:2003ms,Aspinwall:2009ms})
\begin{eqnarray}
\psi^i \in \Gamma(\phi^* T^{1,0}Z)\,, \quad  && \psi^{\bar{\imath}}_z \in \Gamma(K \otimes \phi^* T^{0,1} Z)\,,\nonumber\\
\chi^i_{\bar{z}} \in \Gamma(\overline{K}\otimes \phi^* T^{1,0}Z)\, , \quad&&   \chi^{\bar{\imath}} \in \Gamma(\phi^* T^{0,1}Z)\, .
\end{eqnarray}
This has the effect that $\psi^i,\chi^{\bar{\imath}}$ become scalars on the $2d$ base manifold of the non-linear sigma model. To analyze the cohomology of 
$$Q_A=Q^{-}   +    \overline{Q}^+ \, ,$$
one sets $\varepsilon^-=\overline{\varepsilon}^+=\varepsilon$ and $\overline{\varepsilon}^-=\varepsilon^+=0$ in (\ref{susytrafo1}-\ref{susytrafo6}) such that the supersymmetry variation $\delta$ is given by 
$$\delta=\varepsilon\, Q_A\, .$$
The variations of the scalar quantities become (\ref{susytrafo1},\ref{susytrafo3}) and (\ref{susytrafo2},\ref{susytrafo6})
\begin{eqnarray}
\delta \phi^i&=&i \varepsilon\, \psi^i\,,\quad \delta \psi^i=0\,, \\
\delta \phi^{\bar{\imath}}&=&i \varepsilon\, \chi^{\bar{i}}\,,\quad \delta \chi^{\bar{\imath}}=0\,, 
\end{eqnarray}
which suggests the identification:
\begin{equation}
\psi^i \leftrightarrow du^i\,, \quad \chi^{\bar{\imath}} \leftrightarrow d\bar{u}^{\bar{\imath}} \,, \quad Q_A \leftrightarrow d=\partial+\overline{\partial}\, ,
\end{equation}
where $u^i,\bar{u}^{\bar{\imath}}$ are local coordinates on $Z$ and $d$ is the de Rham operator.
Operators of the topological field theory are associated to points on the manifold and can hence only be made out of $2d$ scalars, these would have the form
\begin{equation}
\omega(\phi)_{i_1,\dots ,i_p,\bar{\imath}_1,\dots,\bar{\imath}_q} \, \psi^{i_1}\dots \psi^{i_p}\chi^{\bar{\imath}_1}\dots \chi^{\bar{\imath}_q}\, ,
\end{equation}
which can then be identified with
\begin{equation}
\omega(u)_{i_1,\dots ,i_p,\bar{\imath}_1,\dots,\bar{\imath}_q} \, du^{i_1}\dots du^{i_p} d\bar{u}^{\bar{\imath}_1}\dots d\bar{u}^{\bar{\imath}_q}\, .
\end{equation}
The operators of the topological theory are in one to one correspondence with the de Rham cohomology of $Z$. Furthermore the action can now be written as:
\begin{equation}
S=i\int_{\Sigma} \left\{ Q_A,\mathcal{D}\right\} -2\pi i\int_{\Sigma} \phi^*(B+i\mathcal{J})\, ,
\end{equation}
with the K\"ahler form $\mathcal{J}= i g_{i\bar{\jmath}}\, du^{i} \wedge d\bar{u}^{\bar{\jmath}}$ and 
\begin{equation}
\mathcal{D}=2\pi g_{i\bar{\jmath}} \left( \psi^{\bar{\jmath}}_z \, \bar{\partial} \phi^i +\partial \phi^{\bar{\jmath}} \,\chi^{i}_{\bar{z}}\right)\,.
\end{equation}

The second term depends only on the class $\beta=\phi_*(\Sigma)\in H_2(Z,\mathbbm{Z})$. The partition and correlation functions can hence be written as:
\begin{equation}\label{instsum}
Z=\int \left[d\phi\, d\psi\, d\chi\,\right] e^{-S}= \sum_{\beta\in H_2} e^{-2\pi i t_{\beta}} \int_{\textrm{fixed} \, \beta} \left[d\phi\, d\psi\, d\chi\,\right] e^{-i \int \left\{ Q_A,\mathcal{D} \right\}}\,.
\end{equation}
By scaling the term in the exponential in the integrand to infinity it is argued that the dominant contributions come from maps such that  $\mathcal{D}=0$. These are holomorphic maps
$$\bar{\partial} \phi=0\,,$$
and it can be shown that these contributions are exact, which is the idea of localization (See \cite{Hori:2003ms,Aspinwall:2009ms} and references therein for more details). In these cases the path integral can be defined as integrals over moduli spaces of holomorphic maps and gives mathematical invariants which characterize the target space $Z$, the Gromov-Witten invariants. Physically the sum over holomorphic maps wrapping various 2-cycles in the target space corresponds to summing over different world-sheet instanton sectors.


\subsubsection{\it Quantum geometry}
It was shown that the operators of the topological field theory obtained from the A-twist are in correspondence with the de Rham cohomology and moreover the $U(1)$ charges are identified with the degrees of the forms. The geometric interpretation of the chiral ring is given by the wedge product of forms or equivalently by the intersection ring of $Z$. However, due to the summation over the all the instanton sectors as in Eq.~(\ref{instsum}), this ring gets modified, giving the quantum cohomology ring, which can be thought of as a quantum mechanical correction to the intersection ring taking into account a counting of holomorphic maps. The introduction of notions from quantum mechanics into geometry is referred to as quantum geometry. Another deviation from the classical understanding of geometry is the notion of the complexified K\"ahler moduli space. Classically the deformations of the K\"ahler form would define the K\"ahler cone. However, the sigma model suggests that the correct quantity to be considered is the complexified K\"ahler class and the enlarged K\"ahler moduli space, which can be understood in terms of different regions, called phases which may relate different CY geometries \cite{Witten:1993yc,Aspinwall:1993nu,Aspinwall:1993xz}.

In the following a short exposition of the deformation subring of the chiral ring will be given, more mathematical details and background can be found in Refs.~\cite{Voisin:1999ms,Cox:1999ms}. This exposition  follows \cite{Batyrev:1993wa}. Quantum cohomology is developed at large volume which refers to the region where the instanton sum in (\ref{instsum}) has a sensible meaning.
Mathematical notions of mirror symmetry in other regions in moduli space are still being developed, see for example \cite{Bouchard:2007nr} and references therein for mirror symmetry involving the orbifold region in moduli space. 
The deformation subring corresponds to the even cohomology of the CY $Z$. The complexified K\"ahler classes are elements in $H^{2}(Z,\mathbbm{C})$. The pairing is given by the wedge product of forms. To recover the chiral ring structure, an interpretation should be given to  following diagram
\begin{equation}
\begin{array}{ccccccc}
H^0(Z,\mathbbm{C})&\xrightarrow{\nabla_A}&H^2(Z,\mathbbm{C})&\xrightarrow{\nabla_A}&H^4(Z,\mathbbm{C})&\xrightarrow{\nabla_A}&H^6(Z,\mathbbm{C})\,.
\end{array}
\end{equation}
To define the connection let $\eta_0$ be the generator of $H^0(Z,\mathbbm{Z})$, $\eta_a\,,\,a=1,\dots,n=h^{1,1}(Z)$ a basis of $H^2(Z,\mathbbm{Z})$, $\chi_b\,,\,b=1,\dots,n$ a basis of $H^4(Z,\mathbbm{Z})$ which is dual with respect to the symplectic pairing $\langle \eta_a,\chi_b\rangle =\delta_{ab}$ and finally let $\chi_0\in H^6(Z,\mathbbm{Z})$ be the dual to $\eta_0$. The connection $\nabla_A$ is now defined by
\begin{eqnarray}
\nabla_A \eta_0 &=& \sum_{a=1}^n \eta_a \otimes \frac{dq_a}{q_a} \, , \nonumber\\
\nabla_A \eta_c &=& \sum_{a,b=1}^n C_{abc} \,\chi_b \otimes \frac{dq_a}{q_a}\, ,\nonumber\\
\nabla_A \chi_b &=& \chi_0 \frac{dq_b}{q_b}\, , \nonumber\\
\nabla_A \chi_0 &=& 0\,.
\end{eqnarray}
The coefficients $C_{abc}$ are power series in $q_1,\dots ,q_n$ defined by rational curves, whose class is given by $\beta\in H_2(Z,\mathbbm{Z})$ 

$$q_{\beta}:=\exp\left( 2\pi i \int_{\beta} B+i\mathcal{J}\right)\, ,$$

\begin{equation}\label{YukawaA}
 C_{abc}=\int_Z  \eta_a\,\wedge \eta_b\, \wedge \eta_c  +\sum_{\beta\in H_2} n_{[\beta]} \langle \beta,\eta_a \rangle \langle \beta,\eta_b \rangle \langle \beta,\eta_c \rangle \frac{q^{[\beta]}}{1-q^{[\beta]}} \, ,
\end{equation}
where $q^{[\beta]}=q^{c_1}\dots q_n^{c_m}$, $c_a=\langle \beta,\eta_a\rangle \,.$ And where
\begin{equation}
 N_{[\beta]}(\eta_a,\eta_b,\eta_c)=n_{[\beta]} \langle \beta,\eta_a \rangle \langle \beta,\eta_b \rangle \langle \beta,\eta_c \rangle\,,
\end{equation}
are rational numbers called the Gromov-Witten invariants. One sees from (\ref{YukawaA}), that the geometric interpretation of the threepoint function involves a classical part given by the triple intersection as well as a quantum part given by the sum over instantons.


\subsection{B-model}
\subsubsection{\it B-twist}
Returning to the B-twist (\ref{twist}), the modification of the bundles of which the fermions are sections becomes
\begin{eqnarray}
\psi^i_z \in \Gamma(K\otimes \phi^* T^{1,0}Z)\,, \quad  && \psi^{\bar{\imath}} \in \Gamma(\phi^* T^{0,1} Z)\,,\nonumber\\
\chi^i_{\bar{z}} \in \Gamma(\overline{K}\otimes \phi^* T^{1,0}Z)\, , \quad&&   \chi^{\bar{\imath}} \in \Gamma(\phi^* T^{0,1}Z)\, .
\end{eqnarray}
The world-sheet scalars are grouped in the following form:
\begin{equation}
\eta^{\bar{\imath}}= \psi^{\bar{\imath}}+ \chi^{\bar{\imath}}\, , \quad \theta_j=g_{\bar{\imath}j}\left( \psi^{\bar{\imath}}- \chi^{\bar{\imath}}\,\right)\,,
\end{equation}
and $\psi^i_z,\chi_z^i$ are considered to be the $(1,0)$ and $(0,1)$ components of a one-form $\rho^i$.
 To analyze the cohomology of 
$$Q_B=Q^{+}   +    \overline{Q}^+ \, ,$$
one sets $\varepsilon^+=\overline{\varepsilon}^+=\varepsilon$ and $\overline{\varepsilon}^+=\varepsilon^+=0$ in (\ref{susytrafo1}-\ref{susytrafo6}) such that the supersymmetry variation $\delta$ is given by 
$$\delta=\varepsilon\, Q_B\, .$$
The supersymmetry variations suggest the identification \cite{Witten:1991zz}
\begin{equation}
\eta^{\bar{\imath}} \leftrightarrow d\bar{u}^{\bar{\imath}}\,, \quad \theta_i \leftrightarrow \frac{\partial}{\partial u^i} \,, \quad Q_B \leftrightarrow \overline{\partial}\, ,
\end{equation}
Operators of the topological field theory made out of $2d$ scalars have the form
\begin{equation}
\omega(\phi)^{i_1,\dots ,i_p}_{\bar{\imath}_1,\dots,\bar{\imath}_q} \, \eta^{\bar{\imath}_1}\dots \eta^{\bar{\imath}_q}\theta_{i_1}\dots \theta_{i_p}\, ,
\end{equation}
which can then be identified with
\begin{equation}
 d\bar{u}^{\bar{\imath}_1}\dots d\bar{u}^{\bar{\imath}_q}\, \omega(u)^{i_1,\dots ,i_p}_{\bar{\imath}_1,\dots,\bar{\imath}_q} \, \frac{\partial}{\partial u^{i_1}}\dots \frac{\partial}{\partial u^{i_p}}\, .
\end{equation}
The operators of the topological theory are in one to one correspondence with $(0,q)$ forms on $Z^*$ valued in $\wedge^{p} TZ^*$. Furthermore, the action can now be written as:
\begin{equation}
S=i\int_{\Sigma} \left\{ Q_B,\mathcal{D}\right\} + U\, ,
\end{equation}
with 
\begin{eqnarray}
\mathcal{D}&=& g_{j\bar{k}} \left( \rho^j_z\,\bar{\partial} \phi^{\bar{k}} + \rho^j_{\bar{z}}\, \partial \phi^{\bar{k}}\right)\, ,\\
U&=& \int_{\Sigma} \left( -\theta_j D\rho^j-\frac{i}{2} R_{j\bar{\jmath} k\bar{k}} \rho^j \wedge \rho^k\,\eta^{\bar{j}} \theta_l\,g^{l\bar{k}} \right)\,.
\end{eqnarray}

Setting $\mathcal{D}=0$ the maps $\phi$ become trivial since $\partial \phi^{\bar{k}}=\bar{\partial} \phi^{\bar{k}}=0$, these maps map the $2d$ world-sheet to a point in $Z^*$. The B-model receives thus no instanton corrections. The correlation functions correspond to classical integrals. Furthermore the unique up to scale holomorphic $(3,0)$ form $\Omega$ of the CY threefold $Z^*$ can be used to transform the $(0,q)$ forms valued in $\wedge^{p} TZ^*$ into $(3-p,q)$ forms. The deformation operators having charge $(1,1)$ get hence identified with elements in $H^{2,1}(Z^*)$, which parameterize the complex structure deformations.

\subsubsection{\it Variation of Hodge structure}
The moduli space on the B-side corresponds to the moduli space of complex structures of the target space $Z^*$.  For a CY threefold, the bundle corresponding to the deformation subring is the middle dimensional cohomology $H^{3}(Z^*,\mathbbm{C})$. This space has a natural splitting once a given complex structure is chosen, i.e., at a specific point in the complex structure moduli space. The split is
\begin{equation} \label{Hodgesplit}
 H^3(Z^*,\mathbbm{C}) \simeq \bigoplus_{p+q=3} H^{p,q}(Z^*)\, .
\end{equation}
This split identifies the unique up to scale holomorphic $(3,0)$ form $\Omega$. It permits furthermore a natural notion of complex conjugation, namely $\overline{H^{p,q}(Y)}=H^{q,p}(Y)$. The term Hodge structure refers to $H^{3}(Z^*,\mathbbm{C})$, together with the split (\ref{Hodgesplit}) and with a lattice given by $H^{3}(Z^*,\mathbbm{Z})$ which generates $H^{3}(Z^*,\mathbbm{C})$ upon tensoring with $\mathbbm{C}$ (See Refs.~\cite{Gross:2001ms,VoisinHodge} and references therein for more details). The split (\ref{Hodgesplit}) does however not vary holomorphically when the complex structure moduli are varied. The non-holomorphicity of this split is at the origin of the holomorphic anomaly equations which will be discussed later. There is however a different split of the bundle which varies holomorphically over the moduli space of complex structures. This split is given by the Hodge filtration $F^{\bullet}(Z^*)=\{ F^{p}(Z^*)\}_{p=0}^3$, where the spaces in brackets are defined by
\begin{equation}\label{filtration}
H^3=F^0 \supset F^1\supset F^2\supset F^3\supset F^4=0\, , \quad F^{p}(Z^*)=\bigoplus_{a\geq p} H^{a,3-a} (Z^*)\, \subset H^3\, .
\end{equation}
To recover the splitting (\ref{Hodgesplit}) one can intersect with the anti-holomorphic filtration
\begin{equation}
 H^{p,q}(Z^*)=F^{p}(Z^*) \cap \overline{F^{q}(Z^*)}\,. 
\end{equation}
Instead of a fixed target space, one can consider a variation family by varying the complex structure of $Z^*$.
In this case the filtration is equipped with a flat connection $\nabla$ which is called the Gauss-Manin connection with the property $\nabla F^p \subset F^{p-1}$, which is called Griffiths transversality. 

This property permits an identification of the derivatives of $\Omega(x)\footnote{$x$ is a local coordinate on the moduli space of complex structures $\mathcal{M}$ used to indicate the dependence of the form type on the patch in moduli space.} \in F^3$ with elements in the lower filtration spaces. The whole filtration can be spanned by taking multiderivatives of the holomorphic $(3,0)$ form. Fourth order derivatives can then again be expressed by the elements of the basis, which is reflected by the fact that periods of $\Omega(x)$ are annihilated by a system of differential equations of fourth order called the Picard-Fuchs (PF) equations. The PF equations capture the variation of Hodge structure which describes the geometric realization on the B-model side of the deformation of the $\mathcal{N}=(2,2)$ superconformal field theory and its chiral ring \cite{Lerche:1991wm}, see also ref \cite{Ceresole:1993qq} for a review. Schematically the variation takes the form:
\begin{equation}
\begin{array}{ccccccc}
F^{3}&\xrightarrow{\nabla_B}&F^{2}&\xrightarrow{\nabla_B}&F^1&\xrightarrow{\nabla_B}&F^0\,.
\end{array}
\end{equation}

The dimensions of the spaces $(F^3,F^2/F^3,F^1/F^2,F^0/F^1)$ are $(1,h^{2,1},h^{2,1},1)$. Elements in these spaces can be obtained by taking derivatives of  $\Omega(x)$ w.r.t the moduli. A section of the filtration can be given by the following vector $w(x)$ which has $2h^{2,1}+2$ components:
\begin{equation}
  \label{filtration}
w(x)= \left(\Omega(x)\, \quad \theta_i\Omega(x) \quad \theta_{*} \theta_i \Omega(x)\, \quad  \theta_{*}^3 \Omega(x)\, \right)^t\, .
\end{equation}
where $\theta_i = x^i \partial_{x^i}, i=1\dots,h^{2,1}(Z^*)$ with $x^i$ some local coordinates and $x^*$ some fixed choice of coordinate such that the $h^{2,1}$ elements $\theta_i \theta_* \Omega(x)$ span $F^1/F^2$. Using $w(x)$ the period matrix can be defined
\begin{equation}
\Pi(x)_{\beta}^{\phantom{\beta} \alpha}= \int_{\gamma^\alpha} w(x)\, ,\quad \gamma^{\alpha}\in H_3(Z^*)\, ,\quad\alpha,\beta=0,\dots,2 h^{2,1}+1\, ,
\end{equation}
the first row of which corresponds to the periods of $\Omega(x)$. The periods are annihilated by the PF operators. 

Furthermore Griffiths transversality together with the symplectic pairing given by integrating the wedge product of two three forms over the CY allows one to define the threepoint functions\footnote{The sign here is a choice of convention.}
\begin{equation}
 C_{ijk} := -\int_Y \Omega \wedge \nabla_i\nabla_j \nabla_k \Omega\, .
\end{equation}

\subsubsection{\it Special/flat coordinates}
\label{sec:flat-coordinates}
A special set of coordinates on the moduli space of complex structures will be discussed which permit an identification with the physical deformations of the underlying theory. These coordinates are defined within special geometry which was developed studying moduli spaces of $\mathcal{N}=2$ theories.\footnote{See Refs. \cite{deWit:1984pk,Cremmer:1984hj,deWit:1984px,Lerche:1989uy,Candelas:1990pi,Candelas:1990rm,Strominger:1990pd,Lerche:1991wm,Ferrara:1991aj,Ceresole:1992su,Ceresole:1993qq,Freed:1999sm}.} 
Choosing a symplectic basis of 3-cycles $A^I,B_J \in H_3(Z^*)$ and a dual basis $\alpha_I,\beta^J$ of $H^3(Z^*)$ such that 
\begin{eqnarray}
&&A^I\cap B_J=\delta^I_J=-B_J\cap A^I\,,\quad  A^I\cap A^J=B_I\cap B_J=0\, \, ,\nonumber\\
&& \int_{A^I} \alpha_J =\delta^I_J\, , \quad\int_{B_J} \beta^I =\delta^I_J\, ,\quad  I,J=0,\dots h^{2,1}(Z^*)\, ,
\end{eqnarray}
the $(3,0)$ form $\Omega(x)$ can be expanded in the basis $\alpha_I,\beta^J$:\footnote{The relative sign choice is a convention which varies in the literature, this choice identifies $X^I=\int_{A^I}\Omega\,,F_I=\int_{B_I}\Omega$.}
\begin{equation}
\Omega(x)= X^I(x) \alpha_I + \mathcal{F}_J(x) \beta^J\, .
\end{equation}
The periods $X^I(x)$ can be identified with projective coordinates on $\mathcal{M}$ and $\mathcal{F}_J$ with derivatives of a homogeneous function $\mathcal{F}(X^I)$ of weight 2 such that $\mathcal{F}_J=\frac{\partial \mathcal{F}(X^I)}{\partial X^J}$. In a patch where $X^0(x)\ne 0$ a set of special coordinates can be defined $$t^a=\frac{X^a}{X^0}\, ,\quad a=1,\dots ,h^{2,1}(Z^*).$$ The normalized holomorphic $(3,0)$ $v_0= (X^0)^{-1} \Omega(t)$ has the expansion:
\begin{equation}
 v_0= \alpha_0 + t^a \alpha_a +\beta^b F_b(t) + (2F_0(t)-t^c F_c(t)) \beta^0\,,
\end{equation}
where $$F_0(t)= (X^0)^{-2} \mathcal{F} \quad \textrm{and} \quad F_a(t):=\partial_a F_0(t)=\frac{\partial F_0(t)}{\partial t^a}.$$
$F_0(t)$ is the prepotential. One can define 
\begin{eqnarray}
v_a &=& \alpha_a+\beta^b F_{ab}(t)+(F_a(t)-t^bF_{ab}(t))\beta^0\, ,\\
v_D^a &=& \beta^a -t^a \beta^0\, ,\\
v^0 &=& -\beta^0\, .
\end{eqnarray}
The Yukawa coupling in special coordinates is given by
\begin{equation}
C_{abc}:= \partial_a \partial_b \partial_c F_0(t)=-\int_{Z^*} v_0 \wedge \partial_a\partial_b\partial_c v_0\, .
\end{equation}
Defining further the vector with $2h^{2,1}+2$ components:
\begin{equation}
v=(v_0\,, \quad v_a\,,\quad v_D^a\,,\quad v^0)^t\,,
\end{equation}
which gives by construction:
\begin{equation}\label{flatconnection}
\partial_a \left( \begin{array}{c} v_0\\v_b\\v_D^b\\v^0 \end{array}\right)=\underbrace{\left( \begin{array}{cccc}
0&\delta^c_a&0&0\\
0&0&C_{abc}&0\\
0&0&0&\delta^b_a\\
0&0&0&0
 \end{array}
\right)}_{:=C_a} \, \left( \begin{array}{c} v_0\\v_c\\v_D^c\\v^0 \end{array}\right)\,,
\end{equation}
which defines the $(2h^{2,1}+2) \times (2h^{2,1}+2)$ matrices $C_a$, in terms of which the equation can be written in the form:
\begin{equation}\label{GaussManin}
\left(\partial_a-C_a\right) \, v=0\, .
\end{equation}

The entries of $v$ correspond to elements in the different filtration spaces discussed earlier.  Eq.~(\ref{GaussManin}) defines the Gauss-Manin connection in special coordinates. The upper triangular structure of the connection matrix reflects the effect of the charge increment of the elements in the chiral ring upon insertion of a marginal operator of unit charge. Since the underlying superconformal field theory is isomorphic for the A- and the B-models, this set of coordinates describing the variation of Hodge structure is the good one for describing mirror symmetry and provide thus the mirror maps.  The power of mirror symmetry comes from the fact that the computation on the B-model side are well under control since they are connected to the variation of complex structure. The computations which are involved are classical. The identification with the non-trivial quantum geometry of the A-model is done by finding the right set of mirror maps, which are the special coordinates which are also called flat coordinates. These are the coordinates in terms of which the Gauss-Manin connection has the form of Eq.~(\ref{flatconnection}). There are several techniques \cite{Lerche:1991wm,Ferrara:1991aj,Ceresole:1992su,Ceresole:1993qq,Hosono:1995bm,Hosono:1996jv,Noguchi:1996tu,Masuda:1998eh,Alim:2012ss} to identify the special set of coordinates which allows an identification with the physical parameters and hence with the A-model side. This will be outlined in the following for the example of the quintic.


\subsection{Example of mirror geometries}
In the following an example of a construction of mirror manifolds as well as finding the appropriate mirror maps will be discussed. The exposition will be rather concise referring to \cite{Candelas:1990rm,Cox:1999ms,Gross:2001ms,Hori:2003ms,Klemm:2005tw} and references therein for further details.  
The quintic $Z$ denotes the CY manifold defined by 
\begin{equation}
Z:=\{P(x)=0\} \subset \mathbb{P}^4\,,
\end{equation} 
where $P$ is a homogeneous polynomial of degree 5 in 5 variables $x_1, \dots ,x_5$. The mirror quintic $Z^*$ can be constructed using the Greene-Plesser construction \cite{Greene:1990ud}. Equivalently it may be constructed using Batyrev's dual polyhedra \cite{Batyrev:1993dm} in the toric geometry language\footnote{For a review of toric geometry see Refs.~\cite{Greene:1996cy,Hori:2003ms}.}.
In the Greene-Plesser construction the family of mirror quintics is the one parameter family of quintics defined by
\begin{equation}\label{GP}
 \{ p(Z^*)=\sum_{i=1}^5 x_i^5-\psi \prod_{i=1}^5 x_i=0 \}  \in \mathbbm{P}^4\,,
\end{equation}
after a $(\mathbbm{Z}_5)^3$ quotient and resolving the singularities.

In the following, the mirror construction following Batyrev will be outlined. The mirror pair of CY 3-folds $(Z,Z^*)$ is given as hypersurfaces in toric ambient spaces $(W,W^*)$. The mirror symmetry construction of Ref.~\cite{Batyrev:1993dm} associates the pair $(Z,Z^*)$ to a pair of integral reflexive polyhedra $(\Delta,\Delta^*)$. 

\subsubsection{\it The A-model geometry}
The polyhedron $\Delta$ is characterized by $k$ relevant integral points $\nu_i$ lying in a hyperplane of distance one from the origin in $\mathbbm{Z}^5$, $\nu_0$ will denote the origin following the conventions of Refs. \cite{Batyrev:1993dm,Hosono:1993qy}.  The $k$ integral points $\nu_i(\Delta)$ of the polyhedron $\Delta$ correspond to homogeneous coordinates $u_i$ on the toric ambient space $W$ and satisfy $n=h^{1,1}(Z)$ linear relations:
\begin{equation}\label{toricrel}
\sum_{i=0}^{k-1} l_i^a \, \nu_i=0\, , \quad a=1,\dots,n\,.
\end{equation}
The integral entries of the vectors $l^a$ for fixed $a$ define the weights $l_i^a$ of the coordinates $u_i$ under the $\mathbbm{C}^*$ actions
$$ u_i \rightarrow (\lambda_a)^{l_i^a} u_i\,, \quad \lambda_a \in \mathbbm{C}^*\,.$$

The $l_i^a$ can also be understood as the $U(1)_a$ charges of the fields of the gauged linear sigma model (GLSM) construction associated with the toric variety \cite{Witten:1993yc}. The toric variety $W$ is defined as $W\simeq (\mathbbm{C}^{k}-\Xi)/(\mathbbm{C}^*)^n$, where $\Xi$ corresponds to an exceptional subset of degenerate orbits. To construct compact hypersurfaces, $W$ is taken to be the total space of the anti-canonical bundle over a compact toric variety. The compact manifold $Z \subset W$ is defined by introducing a superpotential $\mathcal{W}_Z=u_0 p(u_i)$ in the GLSM, where $x_0$ is the coordinate on the fiber and $p(u_i)$ a polynomial in the $u_{{i>0}}$ of degrees $-l_0^a$. At large K\"ahler volumes, the critical locus is at $u_0=p(u_i)=0$ \cite{Witten:1993yc}. 

An example of CY manifold is the quintic which is the compact geometry given by a section of the anti-canonical bundle over  $\mathbbm{P}^4$. The charge vectors for this geometry are given by:
\begin{equation}\label{chargevec}
\begin{array}{ccccccc}
&u_0&u_1&u_2&u_3&u_4&u_5\\
l=&(-5&1&1&1&1&1\, )\,.
\end{array}
\end{equation}
The vertices of the polyhedron $\Delta$ are given by:
\begin{eqnarray}
&&\nu_0=(0,0,0,0,0)\,,\quad \nu_1=(1,0,0,0,0)\,,\quad \nu_2=(0,1,0,0,0)\,,\nonumber\\
&&\nu_3=(0,0,1,0,0)\,,\quad \nu_4=(0,0,0,1,0)\,,\quad \nu_5=(-1,-1,-1,-1,0)\,.
\end{eqnarray}

\subsubsection{\it The B-model geometry}
The B-model geometry $Z^*\subset W^*$ is determined by the mirror symmetry construction of Refs.~\cite{Hori:2000kt,Batyrev:1993dm} as the vanishing locus of the equation
\begin{equation}
p(Z^*)=\sum_{i=0}^{k-1} a_i y_i =\sum_{\nu_i\in \Delta} a_i X^{\nu_i}\, ,
\end{equation}
where $a_i$ parameterize the complex structure of $Z^*$, $y_i$ are homogeneous coordinates \cite{Hori:2000kt} on $W^*$ and $X_m\, , m=1,\dots,4$ are inhomogeneous coordinates on an open torus $(\mathbbm{C}^*)^4 \subset W^*$  and $X^{\nu_i}:=\prod_m X_m^{\nu_{i,m}} $ \cite{Batyrev:1993wa}. The relations (\ref{toricrel}) impose the following relations on the homogeneous coordinates
\begin{equation}
\prod_{i=0}^{k-1} y_i^{l_i^a}=1\, ,\quad a=1,\dots,n=h^{2,1}(Z^*)=h^{1,1}(Z)\, .
\end{equation}
The important quantity in the B-model is the holomorphic $(3,0)$ form which is given by:
\begin{equation}\label{defomega0}
\Omega(a_i)= \textrm{Res}_{p=0} \frac{1}{p(Z^*)} \prod_{i=1}^4 \frac{dX_i}{X_i} \, .
\end{equation}
Its periods 
\begin{equation}
\pi_{\alpha}(a_i)=\int_{\gamma^\alpha} \Omega(a_i)\, , \quad \gamma^{\alpha} \in H_3(Z^*)\,,\quad\alpha=0,\dots, 2h^{2,1}+1\, ,
\end{equation} 
are annihilated by an extended system of GKZ  \cite{Gelfand:1989} differential operators
\begin{eqnarray}
&&\mathcal{L}(l)= \prod_{l_i >0} \left( \frac{\partial}{\partial a_i}\right)^{l_i} -\prod_{l_i<0} \left( \frac{\partial}{\partial a_i}\right)^{-l_i}\, ,\\
&&\mathcal{Z}_k =\sum_{i=0}^{k-1} \nu_{i,j} \theta_i\, , \quad j=1,\dots,4\, . \quad \mathcal{Z}_0 = \sum_{i=0}^{k-1} \theta_i +1\,,\quad \theta_i=a_i \frac{\partial}{\partial a_i}\,,
\end{eqnarray}
where $l$ can be a positive integral linear combination of the charge vectors $l^a$. The equation $\mathcal{L}(l)\, \pi_0(a_i)=0$ follows from the definition (\ref{defomega0}). The equations $\mathcal{Z}_k\,\pi_\alpha(a_i)=0$ express the invariance of the period integral under the torus action and imply that the period integrals only depend on special combinations of the parameters $a_i$
\begin{equation}\label{lcs}
\pi_\alpha(a_i) \sim \pi_\alpha(z_a)\, ,\quad z_a=(-)^{l_0^a} \prod_i a_i^{l_i^a}\, ,
\end{equation}
the $z_a\,, a=1,\dots,n$ define local coordinates on the moduli space $\mathcal{M}$ of complex structures of $Z^*$.

The charge vector defining the A-model geometry in Eq.~(\ref{chargevec}) gives the mirror geometry defined by:
\begin{equation}\label{Batyrev}
 p(Y)=\sum_{i=0}^5 a_i y_i =0\, ,
\end{equation}
where the coordinates $y_i$ are subject to the relation
\begin{equation}
 y_1 y_2 y_3 y_4 y_5 = y_0^5\,.
\end{equation}
Changing the coordinates $y_i=x_i^5,\, i=1,\dots,5$ shows the equivalence of (\ref{GP}) and (\ref{Batyrev}) with 
\begin{equation}
 \psi^{-5}=-\frac{a_1 a_2 a_3 a_4 a_5}{a_0^5}=: z\,.
\end{equation}

Furthermore, the following Picard-Fuchs  (PF) operator annihilating $\tilde{\pi}_{\alpha}(z_i)=a_0\,\pi_{\alpha}(a_i)$ is found:
\begin{equation}\label{PF}
\mathcal{L}=\theta^4- 5z \prod_{i=1}^4 (5\theta+i)\,, \quad \theta=z \frac{d}{dz}\, .
\end{equation}
The discriminant of this operator is
\begin{equation}
  \label{eq:Discriminant}
\Delta=1-3125\,z\,.
\end{equation}
and the Yukawa coupling can be computed:
\begin{equation}
C_{zzz}=\frac{5}{z^3\, \Delta}\,.
\end{equation}

The PF operator gives a differential equation which has three regular singular points which correspond to points in the moduli space of the family of quintics where the defining equation becomes singular or acquires additional symmetries, these are the points \footnote{For further details see \cite{Gross:2001ms}, for the physical SCFT interpretation see \cite{Aspinwall:1994ay}.}
\begin{itemize}
 \item $z=0$ , the quintic at this value corresponds to the quotient of $\prod_{i=1}^5 x_i=0$ which is the most degenerate Calabi-Yau and corresponds to large radius when translated to the A-side.
\item $z=5^{-5}$ this corresponds to a discriminant locus of the differential equation (\ref{PF}) and also to the locus where the Jacobian of the defining equation vanishes. This type of singularity is called a \emph{conifold} singularity. Its A-model interpretation is going to be very useful in the context of higher genus amplitudes in at a later stage.
\item $z=\infty$ , this is known as the orbifold point in the moduli space of the quintic and it corresponds to a non-singular CY threefold with a large automorphism group. This is reflected by a monodromy of order 5.
\end{itemize}

\subsubsection{\it Finding the special/flat coordinates}

The deformations of complex structure of the B-model geometry are typically deformations of the defining equation such as Eq.~(\ref{GP}). The geometric quantities which correspond to the physical generating functions of interest are computed as functions of these deformation parameters. A physical interpretation of the generating functions in terms of instanton sums using the mirror map to the A-model requires finding the good coordinates to describe the deformation problem on the B-model side. These are the special/flat coordinates discussed earlier. It will be sketched in the following how these can be found in the locus in the moduli space of the mirror quintic which is in the vicinity of $z=0$. This is the patch in moduli space mirror to the large volume of the A-model, which is the locus which admits an interpretation in terms of quantum cohomology. The logic is to span the Hodge filtration by acting with derivatives on $\Omega$ and expressing the PF equation as a connection on the deformation bundle. By a change of normalizations and of coordinates, the connection can be brought to the upper triangular form which reflects the chiral ring property. Traditionally, the singularity structure at large volume also permits an identification of the mirror map and the Yukawa couplings as reviewed in \cite{Gross:2001ms} for example. Transforming the connection on the deformation bundle however allows one to find the physical coordinates in various patches in moduli space, which is required for many applications, see for example \cite{Hosono:1996jv,Noguchi:1996tu,Alim:2012ss}.

Choosing the following section of the Hodge filtration:
\begin{equation}
\vec{\Omega}_z(z)=(\Omega(z),\partial_z \Omega(z),\partial_z^2 \Omega(z), \partial_z^3 \Omega(z))^t\,,
\end{equation}
one can use the Picard-Fuchs equation (\ref{PF}) to read off the following connection matrix
\begin{equation}
\partial_z \vec{\Omega}= 
\underbrace{\left(
\begin{array}{cccc}
 0 & 1 & 0 & 0 \\
 0 & 0 & 1 & 0 \\
 0 & 0 & 0 & 1 \\
 \frac{120}{z^3 \Delta} & -\frac{1-15000 z}{z^3\Delta} & -\frac{7-45000 z}{z^2 \Delta} &
   -\frac{6-25000 z}{z \Delta}
\end{array}
\right)}_{M_z} \vec{\Omega}\,,
\end{equation}
where $\Delta=1-3125 z$. This can be written as 
\begin{equation}
 \nabla_z \vec{\Omega}_z=(\partial_z - M_z) \vec{\Omega}_z=0\,.
\end{equation}
To find the basis and the coordinate in terms of which the connection becomes the topological connection, the following ansatz is made
\begin{equation}
\vec{\Omega}_t(z(t))=\left(\frac{\Omega(z(t))}{S_0(t)},\partial_{t} \frac{\Omega(z(t))}{S_0(t)}, C(z(t))^{-1}\partial_{t}^2\frac{\Omega(z(t))}{S_0(t)},\partial_{t} \left(C(z(t))^{-1}\partial_{t}^2\frac{\Omega(z(t))}{S_0(t)}\right) \right)^t \,,
\end{equation}
the vector transforms as
$$ \vec{\Omega}(z(t))_t=A(t) \vec{\Omega}_z \, .$$
Transforming the connection matrix $M_z$ with
$$M_{t}= (\partial_{t} A(t) + \frac{\partial z(t)}{\partial t}A(t) \cdot M_{z}(z(t))) \cdot A^{-1}(t) \,. $$
Requiring the matrix $M_t$ to have the form:
\begin{equation}
M_t= \left(
\begin{array}{cccc}
 0 & 1 & 0 & 0 \\
 0 & 0 & C(z(t)) & 0 \\
 0 & 0 & 0 & 1 \\
 0&0&0&0
\end{array}
\right) \, ,
\end{equation}
gives differential equations for the functions in the ansatz, which are solved by:
\begin{eqnarray*}
S_0(t)&=& 1+120 q + 21000 q^2 +  14115000 q^3+\dots\,,  \quad q=e^{-t}\,,\\ 
z(t)&=& q-770 q^2+171525 q^3+\dots \,, \\
C_{ttt} &=& \partial_t^3 F_0= 5+ 2875 q+4876875 q^2+8564575000 q^3+\dots \, .
\end{eqnarray*}


\section{Higher genus recursion}
\label{highergenus}
The central theme in the following are the holomorphic anomaly equations of Bershadsky, Cecotti, Ooguri and Vafa  (BCOV) \cite{Bershadsky:1993ta,Bershadsky:1993cx}. These relate recursively the topological string amplitudes at genus $g$, $\mathcal{F}^{g}$ which were defined in sec.~(\ref{deftopstring}) to amplitudes of lower genus.  

The derivation of BCOV is based on a worldsheet analysis of the SCFT underlying topological string theory. It was previously shown that one can construct four topological string theories out of the SCFT depending on which finite chiral ring one wants to restrict the attention to. These were the A/\emph{anti}-A-model and the B/anti-B-models corresponding to the $(a,c)/(c,a)$ and $(c,c)/(a,a)$ rings, respectively. The key insight which allows the derivation of the equations is the failure of decoupling of deformations coming from the anti-ring. For instance, deformations coming from the $(a,a)$ ring will have an effect on the amplitudes of the B-model which is supposed to be only deformed by marginal fields coming from the $(c,c)$ ring. The starting point for the analysis are the $tt^*$ equations of Cecotti and Vafa \cite{Cecotti:1991me}. The setup for these is to reconsider the deformation subring discussed previously and to replace half of the states with states coming from worldsheet CPT conjugate operators. In terms of the variation of Hodge structures the interplay of the two theories leads to a consideration of the non-holomorphic variation of the Hodge split instead of the filtration which varies holomorphically as captured by the Picard-Fuchs operators. The set of equations obtained from the variation are familiar in the language of supergravity where they reflect the special geometry of the manifold of the scalars of the vector multiplets. See \cite{Ceresole:1993qq,Ceresole:1995ca} for an overview.

 \subsection{$tt^*$ equations}
To state the $tt^*$ equations, half of the states in the deformation bundle are replaced by states coming from the worldsheet CPT conjugate operators. The states created by the charge $(2,2)$ and $(3,3)$ operators that were introduced with upper indices with respect to the topological metric are replaced using the so called $tt^*$ metric which is obtained using the worldsheet CPT operator $\Theta$ \footnote{The notation here follows \cite{Walcher:2007tp}.},\footnote{Letters from the beginning of the alphabet are running from $0,\dots,n$ while letter from the middle of the alphabet are only running over $1,\dots,n$.}
\begin{equation}
 g_{a\bar{b}}= \langle  e_a|\Theta e^b \rangle \, .
\end{equation}
Using this metric a new basis for the dual states can be defined as follows
\begin{equation}
 |e_{\overline{i}} \rangle = g_{k\overline{i}}|e^k \rangle \,,  \quad |e_{\bar{0}} \rangle = g_{0\bar{0}}|e^0 \rangle \, .
\end{equation}
This $tt^*$ metric induces a connection on the bundle $\mathcal{V}$ which is compatible with the holomorphic structure. It is given by
\begin{equation}
 D_i (| e_a\rangle)=\partial_i (|e_a\rangle) -  (A_i)^b_a (| e_b\rangle) \,,
\end{equation}
where $(A_i)_b^a= g^{b\bar{c}}\partial_i g_{a\bar{c}}$ and $(|e_a\rangle)$ denotes the vector $(|e_0 \rangle,|e_i \rangle,|e_{\bar{\imath}} \rangle,|e_{\bar{0}} \rangle)^T$, in matrix form the connection reads
\begin{equation}
 A_i=\left(\begin{array}{cccc}
       g^{0\bar{0}}\partial_i g_{0\bar{0}} &0&0&0 \\ 0&g^{l\bar{\jmath}}\partial_i g_{k\bar{\jmath}}&0&0 \\ 0&0&0&0 \\0&0&0&0 
      \end{array}
\right) \,,\quad
A_{\bar{\imath}}=\left(\begin{array}{cccc}
       0&0&0&0 \\ 0&0&0&0 \\ 0&0&g^{l\bar{\jmath}}\partial_{\bar{\imath}} g_{k\bar{\jmath}}&0 \\0&0&0&g^{0\bar{0}}\partial_{\bar{\imath}} g_{0\bar{0}}  
      \end{array}
\right) \,.
\end{equation}
Moreover the action of the chiral ring operators $\phi_i,\phi_{\bar{\imath}}$  becomes in this basis
\begin{equation}
\phi_i  \left( \begin{array}{c}
       |e_0 \rangle\\|e_j \rangle\\|e_{\bar{\jmath}} \rangle\\|e_{\bar{0}} \rangle
       \end{array} \right)
=\underbrace{
\left(\begin{array}{cccc}
      0 &\delta_i^k&0&0 \\ 0&0&C_{ijk}g^{k\bar{k}}&0 \\ 0&0&0&g_{i\bar{\jmath}} \\0&0&0&0 
     \end{array}
\right)}_{:=C_i} \left( \begin{array}
        {c}|e_0 \rangle\\|e_k \rangle\\|e_{\bar{k}} \rangle\\|e_{\bar{0}} \rangle
       \end{array} \right) \,, 
\end{equation}
and
\begin{equation}
\phi_{\bar{\imath}}  \left( \begin{array}{c}
       |e_0 \rangle\\|e_j \rangle\\|e_{\bar{\jmath}} \rangle\\|e_{\bar{0}} \rangle
       \end{array} \right)
=\underbrace{
\left(\begin{array}{cccc}
0&0&0&0\\
g_{\bar{\imath}j}&0&0&0\\
0&C_{\bar{\imath}\bar{\jmath}\bar{k}}g^{\bar{k}k}&0&0\\
0&0&\delta^{\bar{k}}_{\bar{\imath}} 
     \end{array}
\right)}_{:=C_{\bar{\imath}}} \left( \begin{array}
        {c}|e_0 \rangle\\|e_k \rangle\\|e_{\bar{k}} \rangle\\|e_{\bar{0}} \rangle
       \end{array} \right) \,. 
\end{equation}

As pointed out previously the ground-state $|e_0\rangle$ of the SCFT is unique up to scale. This means that it is a section of a line bundle $\mathcal{L}$ over the moduli space $\mathcal{M}$. Moreover, as the $|e_i\rangle$ states are created by the operators which also parameterize the deformations of the theory, these can be identified with sections of $\mathcal{L} \otimes T\mathcal{M}$. The reason they are also sections of the line bundle is because they inherit the arbitrary scale from the ground-state. The bundle of ground-states can hence be organized as
\begin{equation} \label{V}
V_{\mathbbm{C}}= \mathcal{L}\oplus \mathcal{L}\otimes T\mathcal{M} \oplus \overline{ \mathcal{L}\otimes T\mathcal{M}}\oplus \overline{\mathcal{L}}.
\end{equation}
The $tt^*$ equations \cite{Cecotti:1991me}\footnote{For a pedagogical derivation see \cite{Hori:2003ms}.} can now be spelled out
\begin{equation}\label{tt*}
\begin{array}{rclrcl}
[D_i,D_j]&=&0,& [D_{\overline{i}}, D_{\overline{j}} ]& =& 0 , \\
\left[D_i,C_j\right]&=&[D_j,C_i],&   \quad [D_{\overline{i}}, C_{\overline{j}} ] &=& [C_{\overline{i}}, D_{\overline{j}}] ,\\ 
\left[D_i,D_{\overline{j}} \right]&=&-[C_i,C_{\overline{j}}]\,.&&&
\end{array}
\end{equation}
The last one of these equations is particularly interesting as it allows to define a modified connection which has vanishing curvature
\begin{equation}\
 \nabla_i=D_i+C_i\, ,\quad \nabla_{\bar{\imath}}=D_{\bar{\imath}}+C_{\bar{\imath}}\, , \quad \left[ \nabla_i,\nabla_{\bar{\imath}} \right]=0\,.
\end{equation}
This connection is the Gauss-Manin connection which was already encountered in the discussion of topological field theory and variation of Hodge structures. 


\subsection{Special geometry of $\mathcal{M}$}

The $tt^*$ metric on $V_{\mathbbm{C}}$ can further be used to define a metric on $\mathcal{M}$ by restricting to the $T\mathcal{M}$ valued part of the metric and its complex conjugate. The metric obtained in this way is the Zamolodchikov metric which is defined by
\begin{equation}
 G_{i\bar{\jmath}} = \frac{g_{i\bar{\jmath}}}{g_{0\bar{0}}}\, .
\end{equation}
Setting further $$g_{0\bar{0}}:=e^{-K}\,,$$
it is shown that $G_{i\bar{\jmath}}=\partial_i \partial_{\bar{\jmath}}K$ and hence the Zamolodchikov metric is K\"ahler. Furthermore, the last equation of (\ref{tt*}) can be translated into the following statement about the curvature of $G_{i\bar{\jmath}}$ 
\begin{eqnarray}
R_{i\overline{i}\phantom{l}j}^{\phantom{i\overline{i}}l}=[\bar{\partial}_{\overline{i}},D_i]^l_{\phantom{l}j}=\bar{\partial}_{\bar{\imath}} \Gamma^l_{ij}= \delta_i^l
G_{j\bar{\imath}} + \delta_j^l G_{i\bar{\imath}} - C_{ijk} C^{kl}_{\bar{\imath}},
\label{curvature}
\end{eqnarray}
where $D$ is the covariant derivative with connection parts which follow from the context
$$\Gamma^k_{ij} =G^{k\overline{k}} \p_i G_{\overline{k} j},  \quad\textrm{and} \quad K_i=\p_i K,$$
for the cotangent bundle and the line bundle respectively and
$$\overline{C}_{\bar{\imath}}^{jk}:= e^{2K} G^{k\bar{k}} G^{l\bar{l}}\overline{C}_{\bar{\imath}\bar{k}\bar{l}}. $$
The K\"ahler condition together with the condition on the curvature makes the manifold $\mathcal{M}$ a special K\"ahler manifold, the geometry of which is called special geometry. (See \cite{Strominger:1990pd,Ceresole:1993qq,Ceresole:1995ca,Freed:1999sm} and references therein.) The Yukawa coupling in this setting is a holomorphic section of $\mathcal{L}^{2} \otimes \textrm{Sym}^3 T^* \mathcal{M}$.

Here it will be shown that the $tt^*$ formulation of the geometry of the deformation bundle corresponds on the B-model side to the non-holomorphic variation of the Hodge bundle
\begin{equation}  
V_{\mathbbm{C}}= H^3(X,\mathbbm{C})=H^{3,0}(X)\oplus H^{2,1}(X)\oplus H^{1,2}(X)\oplus H^{0,3}(X)\, .
\end{equation}
Taking as  section of $\mathcal{L}$ the unique up to scale holomorphic $(3,0)$ form $\Omega \in H^{3,0}(X)$ and the inner product which is given by taking wedge products of forms and integrating over $X$, the metric on the line bundle is 
\begin{equation}
 g_{0\bar{0}}=e^{-K}=-i\int_X \Omega \wedge \overline{\Omega}\,.
\end{equation}
One can define elements in $H^{2,1}$ and $H^{1,2}$ in the following way
\begin{equation}
 \chi_i:=D_i \Omega= \left(\partial_i+K_i \right)\Omega\,,\quad  \overline{\chi}_{\bar{\imath}}:=D_{\bar{\imath}} \overline{\Omega}= \left(\partial_{\bar{\imath}}+K_{\bar{\imath}}\right) \overline{\Omega}\,.
\end{equation}
With inner product
\begin{equation}
 g_{i\bar{\jmath}}=i\int_{X} \chi_i \wedge \overline{\chi_{\bar{\jmath}}}=e^{-K} G_{i\bar{\jmath}}=e^{-K} \partial_i \partial_{\bar{\jmath}} K\,.
\end{equation}
Defining the Yukawa coupling
\begin{equation}
 C_{ijk}:= -\int_X \Omega \wedge D_i D_j D_k \Omega = -\int_X \Omega \wedge \partial_i \partial_j \partial_k \Omega\,,
\end{equation}
one can compute
\begin{equation}
 D_i D_j \Omega= D_i\chi_j= i C_{ijk}e^K G^{k\bar{k}} \overline{\chi_{\bar{k}}}\,, \quad D_i \overline{\chi}_{\bar{\imath}}=G_{i\bar{\imath}} \overline{\Omega}\,.
\end{equation}
Taking as a section of $V_{\mathbbm{C}}$ the following vector
\begin{equation}
 \vec{\Omega}=(\Omega,\chi_i,\overline{\chi_{\bar{\imath}}},\overline{\Omega})\, ,
\end{equation}
one recovers the Gauss-Manin connection

\begin{equation}
D_i  \left( \begin{array}{c}
       \Omega\\\chi_j\\\overline{\chi_{\bar{\jmath}}}\\\overline{\Omega}
       \end{array} \right)
=
\left(\begin{array}{cccc}
      0 &\delta_i^k&0&0 \\ 0&0&iC_{ijk}e^K G^{k\bar{k}}&0 \\ 0&0&0&G_{i\bar{\jmath}} \\0&0&0&0 
     \end{array}
\right) \left( \begin{array}
        {c} \Omega\\\chi_k\\\overline{\chi_{\bar{k}}}\\\overline{\Omega}
       \end{array} \right) \,. 
\end{equation}
This matrix equation can be translated into the following fourth order differential equation for $\Omega$
\begin{equation}
 D_m D_n (C_i^{-1})^{lj} D_i D_j \Omega=0\,,
\end{equation}
which is the Picard-Fuchs equation. In Ref.~\cite{Ceresole:1992su,Ceresole:1993qq} it is shown that this equation is equivalent to the holomorphic Picard-Fuchs equations encountered previously and that choosing the non-holomorphic variation of Hodge structures is merely a choice of gauge for the vector $\vec{\Omega}$. 


\subsection{Anomaly equations and their recursive solution}
It was argued that deformations of the topological string theory corresponds to the addition of terms of the form
$$ z^i \int \phi_i^{(2)}+ \bar{z}^{\overline{i}} \int \bar{\phi}_{\overline{i}}^{(2)},$$
to the action $S$. $z^i$ are complex coordinates on $\mathcal{M}$, the notation change from $t^i$ to $z^i$ is to make manifest the difference between the topological/special coordinates $t^i$ for which taking derivatives corresponds to insertions and the more general coordinates $z^i$ for which insertions correspond to taking covariant derivatives. Hence, amplitudes with insertions of $n$ descendants of marginal (c,c) fields (for the B-model) can be defined by\ \cite{Bershadsky:1993cx}
\begin{equation}
 \mathcal{F}^{(g)}_{i_1\dots i_n} = D_{i_n} \mathcal{F}^{(g)}_{i_1\dots i_{n-1}}\,.
\end{equation}
Taking a derivative with respect to $\overline{z}^{\bar{\imath}}$ on the other hand corresponds to inserting a descendant of marginal $(a,a)$ operator
\begin{equation}
\phi_{\bar{\imath}}^{(2)}=dz d\bar{z} \{ G^+,[ \overline{G}^+,\phi_{\bar{\imath}}]\}= -\frac{1}{2} dz d\overline{z} \{ G^+ +\overline{G}^{+},[ G^+-\overline{G}^{+},\phi_{\bar{\imath}}]\}\,.
\end{equation}
Within topological \emph{field} theory such an insertion in a correlator would vanish as the second part of the equation shows that the operator is 
$Q_B$ exact. The field theory argument for this decoupling is that the operator $Q_B$ is a symmetry of the theory and can be commuted through the correlator until it acts on a ground-state and gives zero. The failure of this argument in the case of topological string theory is outlined in the following. The formula for $\mathcal{F}^g$ is given again
\begin{equation}
 \mathcal{F}^g= \int_{\mathcal{M}_g} [dm \,d\bar{m}] \langle \prod_{a=1}^{3g-3}   (\int_{\Sigma} \mu_a G^-) (\int_{\Sigma} \mu_{\overline{a}} \overline{G}^-)\rangle_{\Sigma_g}\,,
\end{equation}
one can see that while commuting $G^+$ and $\overline{G}^+$ through the correlation function defining $\mathcal{F}^g$ that these will hit the $G^-$ and $\overline{G}^-$ which are contracted with the Beltrami differentials, this will give by the superconformal algebra the energy momentum tensor $T$. The energy momentum tensor in turn measures the response of the theory to a variation of the metric and can hence be traded for derivatives $\frac{\partial^2}{\partial m \partial \overline{m}}$on the moduli space of Riemann surfaces $\mathcal{M}_g$ as this latter space precisely parameterizes such changes. One is thus finally left with an integral of a derivative with respect to the moduli of Riemann surfaces over the moduli space of Riemann surfaces which is zero up to boundary contributions. It is exactly these contributions coming from the boundary of the moduli space of Riemann surfaces which is arranged in the holomorphic anomaly equations. These are pictured diagrammatically in the following, for the full derivation of the equations reference is given to the original work \cite{Bershadsky:1993ta,Bershadsky:1993cx}. The equations read
\begin{eqnarray} \label{anom1}
\bar{\partial}_{\bar{\imath}} \mathcal{F}^{(g)} = \frac{1}{2} \bar{C}_{\bar{\imath}}^{jk} \left(
\sum_{g_1+g_2=g}
D_j\mathcal{F}^{(g_1)} D_k\mathcal{F}^{(g_2)} +
D_jD_k\mathcal{F}^{(g-1)} \right) \,,
\end{eqnarray}
\begin{center}
\includegraphics[width=0.7\textwidth]{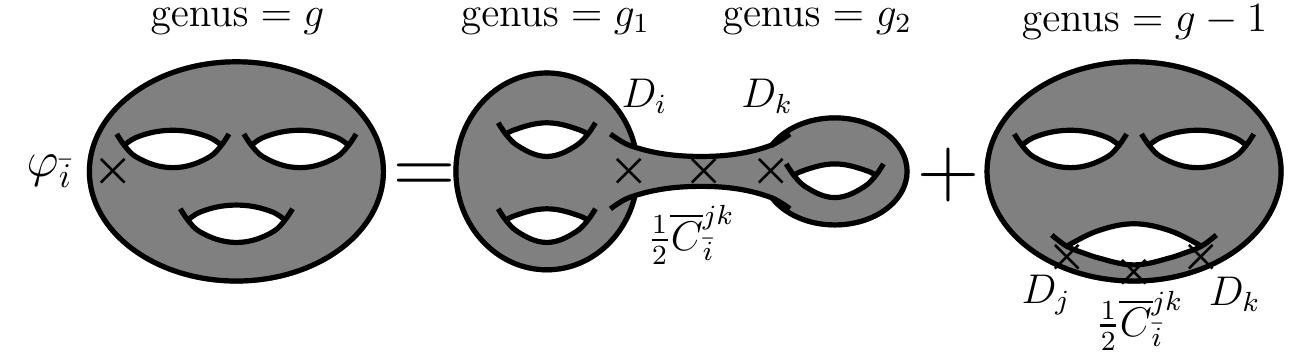}
\end{center}
where $g>1$ and
\begin{eqnarray}
\bar{C}_{\bar{k}}^{ij}= \bar{C}_{\bar{\imath} \bar{\jmath}\bar{k}} G^{i
\bar{\imath}}G^{j \bar{\jmath}}\, \textrm{e}^{2K}, \qquad \bar{C}_{\bar{\imath}\bar{\jmath}\bar{k}}=
\overline{C_{ijk}}. 
\end{eqnarray}
For $g=1$ the equation is
\begin{eqnarray}\label{anom2}
\bar{\partial}_{\bar{\imath}} \mathcal{F}^{(1)}_j &=& \frac{1}{2} C_{jkl}
C^{kl}_{\bar{\imath}}+ (1-\frac{\chi}{24})
G_{j \bar{\imath}}\,.
\end{eqnarray}

\subsubsection*{\it Solution in terms of Feynman diagrams}
BCOV proposed together with the anomaly equation \cite{Bershadsky:1993cx} a method to recursively solve for the $\mathcal{F}^{(g)}$ s. The idea is to partially integrate the anti-holomorphic derivative appearing in the anomaly equation until one ends up with an expression of the form
 $$ \p_{\overline{i}}A= \p_{\overline{i}} B \\
\Rightarrow  A=B + f(z),$$
where $f(z)$ is the holomorphic ambiguity. To be able to do this in practice they first note that the anti-holomorphic Yukawa coupling can locally be written as
$$ \overline{C}_{\overline{i}\overline{j}\overline{k}}= e^{-2K} D_{\overline{i}}D_{\overline{j}} \overline{\p}_{\overline{k}} S, $$
where $S$ is a section of $\mathcal{L}^{-2}$. One can further introduce some objects $S^{ij},S^{i}$ which are sections of $\mathcal{L}^{-2}\otimes \textrm{Sym}^2 (T^*\mathcal{M})$ and of   $\mathcal{L}^{-2}\otimes T^*\textrm(\mathcal{M}) $    such that
\begin{eqnarray}\label{prop}
\partial_{\bar{\imath}} S^{ij}= \bar{C}_{\bar{\imath}}^{ij}, \qquad
\partial_{\bar{\imath}} S^j = G_{i\bar{\imath}} S^{ij}, \qquad
\partial_{\bar{\imath}} S = G_{i \bar{\imath}} S^i.
\end{eqnarray}
Now the procedure is to successively partially integrate the anomaly equation as follows, first write
\begin{eqnarray} \bar{\partial}_{\bar{\imath}} \big( \mathcal{F}^{(g)} - \frac{1}{2}  S^{jk}(
\sum_{{g_1+g_2=g}}
D_j\mathcal{F}^{(g_1)} D_k\mathcal{F}^{(g_2)} + 
 D_jD_k\mathcal{F}^{(g-1)})\big) \\\nonumber
= - \frac{1}{2} S^{jk}\bar{\partial}_{\bar{\imath}}(
\sum_{{g_1+g_2=g}}
D_j\mathcal{F}^{(g_1)} D_k\mathcal{F}^{(g_2)} + 
 D_jD_k\mathcal{F}^{(g-1)}\big),
\end{eqnarray}

then the repeated use of the expression found for the curvature $[\overline{\p}_{\overline{i}},D_j]$ and the knowledge of $\overline{\p}_{\overline{i}} \mathcal{F}^{\tilde{g}}$ with $\tilde{g}< g$ yields the desired result. They further remarked that the expressions found this way resemble Feynman diagrams if one considers a quantum system with one more degree of freedom than the dimension of the moduli space $\mathcal{M}$, where the propagators are given by
$$ K^{ij}=-S^{ij}, \qquad K^{i \phi}= -S^i ,\quad  \textrm{and} \quad K^{\phi \phi}=- 2 S, $$
and vertices are given by the correlation functions with insertions. Furthermore a proof was given for the expansion in terms of Feynman diagrams. \\
One of the practical shortcomings of this procedure is however that the number of iterations grows very fast, expressions for the $\mathcal{F}^{(g)}$ become very long. On the other hand the recursive information contained in the anomaly equation needs to be supplemented by boundary conditions at every step in order to fix the holomorphic ambiguities.


\subsection{Polynomial structure of higher genus amplitudes}\label{sec:polynomial}
The higher genus free energies of the topological string can be interpreted as giving certain amplitudes of the physical string theory.\footnote{See Ref.~\cite{Antoniadis:2007ta} for a review.} The full topological string partition function conjecturally also encodes the information of $4d$ BPS states \cite{Ooguri:2004zv}. It is thus natural to expect the topological string free energies to be characterized by automorphic forms of the target space duality group. The modularity of the topological string amplitudes was used in \cite{Bershadsky:1993cx} to fix the solutions of the anomaly equation. The modularity of the amplitudes is most manifest whenever the modular group is $\mathrm{S}\mathrm{L}(2,\mathbbm{Z})$ or a subgroup thereof. The higher genus generating functions of the Gromov-Witten invariants for the elliptic curve were expressed as polynomials \cite{Rudd:1994ta,Dijkgraaf:1995} where the polynomial generators were the elements of the ring of almost holomorphic modular forms $E_2,E_4$ and $E_6$ \cite{Kaneko:1995}. Polynomials of these generators also appear whenever $\mathrm{S}\mathrm{L}(2,\mathbbm{Z})$ is a subgroup of the modular group, as for example in Refs.~\cite{Minahan:1998vr,Hosono:1999qc,Hosono:2002xj,Klemm:2004km,Alim:2012ss,Klemm:2012sx}. The relation of topological strings and almost holomorphic modular forms was further explored in Refs.~\cite{Aganagic:2006wq,Gunaydin:2006bz,Grimm:2007tm}. For most moduli spaces of CY manifolds and their monodromy groups, the appropriate automorphic forms are however not known. It turns out that using special geometry is enough to prove a polynomial structure of the higher genus topological string amplitudes \cite{Yamaguchi:2004bt, Alim:2007qj}, which will be reviewed in the following.

Yamaguchi and Yau proved in \cite{Yamaguchi:2004bt} that the higher genus amplitudes of the topological string for the mirror quintic and related geometries with a one dimensional space of deformations of complex structures can be expressed as polynomials of degree $3g-3+n$ where $n$ refers to the number of insertions, in a finite number of generators. As generators of the polynomials some elements of special geometry were used which appear on the r.h.s of the holomorphic anomaly equation, namely multi-derivatives of the connections. The non-holomorphic generators of \cite{Yamaguchi:2004bt} consist of
$$A_p =G^{z\bar{z}} (z\p_z)^p G_{z\bar{z}} \quad \textrm{and} \quad B_p = e^K (z\p_z)^p e^{-K}, \qquad p=1,2,3,\dots$$
As a holomorphic generator  $X\sim \frac{1}{\Delta}$ with $\Delta$ discriminant was used. In a next step, relations between the generators are shown such that the infinite number of non-holomorphic generators can be reduced to $A_1, B_1, B_2$ and $B_3$. Furthermore from the analysis of the holomorphic anomaly equation it can be proven \cite{Yamaguchi:2004bt} that only special combinations of the generators appear in the topological string amplitudes and thus the number of non-holomorphic generators gets reduced by one.

The generalization of this construction to CY manifolds with a moduli space of arbitrary dimension was given in Ref.~\cite{Alim:2007qj}.\footnote{This was solved independently by P. Mayr in an unpublished manuscript.} This construction will be reviewed in the following along with the necessary boundary conditions to solve the holomorphic ambiguities at each step of the recursion.

In Ref.~\cite{Alim:2007qj} it was proven that the correlation functions
$\mathcal{F}^{(g)}_{i_1\cdots i_n}$ are polynomials of degree $3g-3+n$ in the
generators $K_i,S^{ij},S^{i},S$ where a grading $1,1,2,3$ was assigned to these generators, respectively. It was furthermore shown that by making a change of generators \cite{Alim:2007qj}
\begin{eqnarray}\label{shift}
\tilde{S}^{ij} &=& S^{ij}, \nonumber \\
\tilde{S}^i &=& S^i - S^{ij} K_j, \nonumber \\
\tilde{S} &=& S- S^i K_i + \frac{1}{2} S^{ij} K_i K_j, \nonumber\\
\tilde{K}_i&=& K_i\, , 
\end{eqnarray}
the $\mathcal{F}^{(g)}$ do not depend on $\tilde{K}_i$, i.e., $\partial \mathcal{F}^{(g)}/\partial \tilde{K}_i=0$. The tilde will be dropped from the modified generators in the following.

The proof relies on expressing the first non-vanishing correlation functions in terms of these generators. At genus zero these are
the holomorphic three-point couplings $\mathcal{F}^{(0)}_{ijk} = C_{ijk}$.
The holomorphic anomaly equation Eq.~(\ref{anom1}) can be integrated
using Eq.~(\ref{prop}) to
\begin{equation}
\mathcal{F}^{(1)}_i = \frac{1}{2} C_{ijk} S^{jk} +(1-\frac{\chi}{24}) K_i
+ f_i^{(1)}, \label{sol2}
\end{equation}
with ambiguity $f_i^{(1)}$. As can be seen from
this expression, the non-holomorphicity of the correlation functions
only comes from the generators. Furthermore the special geometry relation (\ref{curvature}) can be integrated:
\begin{equation}
\Gamma^l_{ij} = \delta_i^l K_j + \delta^l_j K_i - C_{ijk} S^{kl} + s^l_{ij}\,,
\label{specgeom}
\end{equation}
where $s^l_{ij}$ denote holomorphic functions that are not fixed by the
special geometry relation, this can be used to derive the following equations which show the closure of the generators carrying the non-holomorphicity under taking derivatives \cite{Alim:2007qj}.\footnote{These equations are for the tilded generators of Eq.~(\ref{shift}) and are obtained straightforwardly from the equations in Ref.~\cite{Alim:2007qj}}
\begin{eqnarray} \label{rel}
\partial_i S^{jk} &=& C_{imn} S^{mj} S^{nk} + \delta_i^j S^k +\delta_i^k S^j-s_{im}^j S^{mk} -s_{im}^k S^{mj} + h_i^{jk} \, , \nonumber\\
\partial_i S^j &=& C_{imn} S^{mj} S^n + 2 \delta_i^j S -s_{im}^j S^m -h_{ik} S^{kj} +h_i^j \, ,\nonumber\\
\partial_i S &=& \frac{1}{2} C_{imn} S^m S^n -h_{ij} S^j +h_i \, ,\nonumber\\
\partial_i K_j &=& K_i K_j -C_{ijn}S^{mn} K_m + s_{ij}^m K_m -C_{ijk} S^k + h_{ij} \, ,
\end{eqnarray}
where $h_i^{jk}, h^j_i$, $h_i$ and $h_{ij}$ denote
holomorphic functions. All these functions together with the functions in Eq.~(\ref{specgeom}) are not independent. It was shown in Ref. \cite{Alim:2008kp} (See also \cite{Hosono:2008ve}) that the freedom of choosing the holomorphic functions in this ring reduces to functions $\mathcal{E}^{ij},\mathcal{E}^j,\mathcal{E}$ which can be added to the polynomial generators
\begin{eqnarray} \label{freedom}
\widehat{S}^{ij} &=& S^{ij} + \mathcal{E}^{ij} \, ,\nonumber\\
\widehat{S}^{j} &=& S^{j} + \mathcal{E}^{j} \, ,\nonumber\\
\widehat{S} &=& S + \mathcal{E} \, .
\end{eqnarray}
All the holomorphic quantities change accordingly.

The topological string amplitudes now satisfy the holomorphic anomaly equations where the $\bar{\partial}_{\bar{\imath}}$ derivative is replaced by derivatives with respect to the polynomial generators \cite{Alim:2012ss,Alim:2007qj}. 
\begin{eqnarray}\label{polrec1}
\bar{\partial}_{\bar{\imath}} \mathcal{F}^{(g)} &=&  \bar{C}_{\bar{\imath}}^{jk} \Big\{  \frac{\partial \mathcal{F}^{(g)}}{\partial S^{jk}} -\frac{1}{2}  \frac{\partial \mathcal{F}^{(g)}}{\partial S^{k}}K_j -\frac{1}{2}  \frac{\partial \mathcal{F}^{(g)}}{\partial S^{j}}K_k + \frac{1}{2}  \frac{\partial \mathcal{F}^{(g)}}{\partial S}K_jK_k\Big\}+G_{\bar{\imath}j} \frac{\partial \mathcal{F}^{(g)}}{\partial K_{j}} \\
&=& \frac{1}{2} \bar{C}_{\bar{\imath}}^{jk} \Big\{\Big(  \partial_j \partial_k +(C_{jkm} S^{ml} -s_{jk}^l) \partial_l   + (4-2g) (h_{jk} -C_{jkl} S^l) \nonumber\\ \label{polrec2}
&&+(3-2g)K_j \,\partial_k+(3-2g)K_k \,\partial_j +(4-2g)(3-2g)K_k K_j  \Big) \mathcal{F}^{(g-1)} +\nonumber\\ 
&&+ \sum_{h=1}^{g-1} \partial_j \mathcal{F}^{(g-h)} \partial_k \mathcal{F}^{(h)} +(2-2h)K_j \mathcal{F}^{(h)} \partial_k \mathcal{F}^{(g-h)}  +(2-2g+2h)K_k \mathcal{F}^{(g-h)} \partial_j \mathcal{F}^{(h)} \nonumber\\ 
&&+K_j K_k (2-2h)(2-2g+2h)\mathcal{F}^{(h)}\,\mathcal{F}^{(g-h)} \Big\} \, .
\end{eqnarray}

\subsection{Using the polynomial structure}

\subsubsection{\it Constructing the generators} \label{generators}
The construction of the generators of the polynomial construction has been discussed in Ref.~\cite{Alim:2008kp}. The starting point is to pick a local coordinate $z_*$ on the moduli space such that $C_{*ij}$ is an invertible $n \times n$ matrix in order to rewrite Eq.~(\ref{specgeom}) as
\begin{equation}\label{schoice}
S^{ij}=(C^{-1}_{*})^{ik} \left( \delta_*^j K_k +\delta_k^j K_{*} -\Gamma_{*k}^j +s_{*k}^j\right)\,,
\end{equation}
The freedom in Eq.~(\ref{freedom}) can be used to choose some of the $s_{ij}^k$ \cite{Alim:2008kp}. The other generators are then constructed using the equations (\ref{rel}) \cite{Alim:2008kp}:
\begin{eqnarray}\label{holchoice}
S^i &=& \frac{1}{2} \left( \partial_i S^{ii}  -C_{imn} S^{mi} S^{ni} + 2 s_{im}^i S^{mi} -h_i^{ii}\right) \,, \\
S&=& \frac{1}{2} \left(  \partial_i S^i -C_{imn} S^m S^{ni} +s_{im}^i S^m + h_{im} S^{mi} -h_i^i\right) \, .
\end{eqnarray}
In both equations the index $i$ is fixed, i.e., there is no summation over that index. The freedom in adding holomorphic functions to the generators of Eq.~(\ref{freedom}) can again be used to make some choice for the functions $h^{ii}_i,h^i_i$, the other ones are fixed by this choice and can be computed from Eq.~(\ref{rel}).


\subsubsection{\it Boundary conditions}

\subsubsection*{\it Genus 1}
The holomorphic anomaly equation at genus $1$  (\ref{anom2}) can be integrated to give:
\begin{equation}
\label{genus1}
\mathcal{F}^{(1)}= \frac{1}{2} \left( 3+n -\frac{\chi}{12}\right) K +\frac{1}{2} \log \det G^{-1} +\sum_i s_i \log z_i + \sum_a r_a \log  \Delta_a \,,
\end{equation}
where $i=1,\dots,n=h^{1,1}(Z)=h^{2,1}(Z^*),$ and $a$ runs over the number of discriminant components. The coefficients $s_i$ and $r_a$ are fixed by the leading singular behavior of $\mathcal{F}^{(1)}$ which is given by \cite{Bershadsky:1993cx}
\begin{equation}
\mathcal{F}^{(1)} \sim -\frac{1}{24} \sum_i \log z_i \int_Z c_2 J_i \, ,
\end{equation}
for the algebraic coordinates $z_i$, for a discriminant $\Delta$ corresponding to a conifold singularity the leading behavior is given by 
\begin{equation} 
 \mathcal{F}^{(1)} \sim -\frac{1}{12} \log \Delta \,.
 \end{equation}

\subsubsection*{\it Higher genus boundary conditions}
The holomorphic ambiguity needed to reconstruct the full topological string amplitudes can be fixed by imposing various boundary conditions for $\mathcal{F}^{(g)}$ at the boundary of the moduli space.

\subsubsection*{{\it The large complex structure limit}}
The leading behavior of $\mathcal{F}^{(g)}$ at this point (which is mirror to the large volume limit) was computed in \cite{Bershadsky:1993ta,Bershadsky:1993cx,Marino:1998pg,Gopakumar:1998ii,Faber:1998,Gopakumar:1998jq}. In particular the 
contribution from constant maps is 
\begin{equation} \label{constmaps}
 \mathcal{F}^{(g)}|_{q_a=0}= (-1)^g \frac{\chi}{2} \frac{|B_{2g} B_{2g-2}|}{2g\,(2g-2)\,(2g-2)!} \; , \quad g>1,
\end{equation}
where $q_a$ denote the exponentiated mirror map at this point.  

\subsubsection*{{\it Conifold-like loci}}

The leading singular behavior of the partition function $\mathcal{F}^{(g)}$ at a conifold locus has been determined in \cite{Bershadsky:1993ta,Bershadsky:1993cx,Ghoshal:1995wm,Antoniadis:1995zn,Gopakumar:1998ii,Gopakumar:1998jq}
\begin{equation} \label{Gap}
 \mathcal{F}^{(g)}(t_c)= b \frac{B_{2g}}{2g (2g-2) t_c^{2g-2}} + O(t^0_c),
\qquad g>1\,,
\end{equation}
Here $t_c\sim \Delta^{\frac{1}{m}}$ is the flat coordinate 
at the discriminant locus $\Delta=0$. For a conifold singularity $b=1$ and $m=1$. In particular the leading singularity in \eqref{Gap} as well as the absence of subleading singular terms follows from the Schwinger loop computation of \cite{Gopakumar:1998ii,Gopakumar:1998jq}, which computes the effect of the extra massless hypermultiplet 
in the space-time theory \cite{Vafa:1995ta}. The singular structure and the ``gap''  
of subleading singular terms have been also observed in the dual matrix model
\cite{Aganagic:2002wv} and were first used in \cite{Huang:2006si,Huang:2006hq} 
to fix the holomorphic ambiguity at higher genus. The space-time derivation of \cite{Gopakumar:1998ii,Gopakumar:1998jq} is
not restricted to the conifold case and applies also to the case $m>1$ 
singularities which give rise to a different spectrum of
extra massless vector and hypermultiplets in space-time. 
The coefficient of the Schwinger loop integral is a weighted trace over the spin of the particles~\cite{Vafa:1995ta, Antoniadis:1995zn} leading to the prediction $b=n_H-n_V$ for the coefficient of the leading singular term. The appearance of the prefactor $b$ in the case $m>1$ has been discussed in~\cite{Alim:2008kp} for the case of the local $\mathbbm{F}_2$ (see also~\cite{Haghighat:2009nr}).

\subsubsection*{{\it The holomorphic ambiguity}}
\label{sec:it-holom-ambig}

\def\tz{\tilde{z}}
The singular behavior of $\mathcal{F}^{(g)}$ is taken into account by the local ansatz
\begin{equation}\label{ansatzha}
\mathrm{hol. ambiguity}\sim \frac{p(\tz_i)}{\Delta^{(2g-2)}},
\end{equation}
for the holomorphic ambiguity near $\Delta=0$, 
where $p(\tz_i)$ is a priori a series in the local coordinates $\tz_i$ near the singularity. Patching together the local information at all the singularities with the boundary 
divisors with finite monodromy, it follows however that the numerator 
$p(z_i)$ is generically a polynomial of low degree in the $z_i$. Here $z_i$ denote the natural coordinates centered at 
large complex structure, $z_i=0\ \forall i$.
The finite number of coefficients in $p(z_i)$ is constrained by \eqref{Gap}. 
The polynomial structure of higher genus amplitudes in combination with the boundary conditions are enough to solve topological string theory perturbatively on local CY manifolds (See for example Refs.~\cite{Haghighat:2008gw,Alim:2008kp}) and they permit progress in solving for higher genus amplitudes on compact CY manifolds \cite{Huang:2006hq,Haghighat:2009nr,Alim:2012ss}.


\subsubsection{\it Example: the quintic}
\subsubsection*{\it Choice of generators}
The choice of the polynomial generators is fixed such that the holomorphic functions appearing in Eqs.~(\ref{specgeom},\ref{rel}) are rational expressions in terms of $z$, the coordinate in the large complex structure patch of the moduli space. For the holomorphic functions in the following lower/upper indices are multiplied/divided by $z$  
$$ A_i^j \rightarrow \frac{z_i}{z_j} A_i^j\,.$$
With this convention, all the holomorphic functions appearing in the setup of the polynomial construction can be expressed in terms of polynomials in the local coordinates (See Refs.~\cite{Yamaguchi:2004bt,Alim:2007qj,Hosono:2008ve}). The holomorphic functions are chosen as in Ref.~\cite{Hosono:2008ve}:
\begin{equation}
s_{zz}^z=-\frac{8}{5}\,, \quad h_{z}^{zz}=\frac{1}{25}\,,\quad h^z_z=-\frac{1}{125}\,,\quad h_z=\frac{2}{3125}\,,\quad h_{zz}=\frac{2}{25}\,,
\end{equation}

\subsubsection*{\it Genus 1}
Using the boundary conditions for genus $1$, the equation for $\mathcal{F}^{1}$ (\ref{genus1}) can be fixed with:
\begin{equation}
n=1\,,\quad \chi=-200\,,\quad  s=-\frac{31}{12}\,, \quad r=-\frac{1}{12}\,,
\end{equation} 
and the starting polynomial expression is fixed to be (with $P=\Delta^{-1}$):
\begin{equation}
z\partial_z\,\mathcal{F}^{(1)}=\frac{1}{60} \left(-112+560 K_z + 5 P (1+30 S^{zz})\right)\,.
\end{equation}

\subsubsection*{\it Higher genus}
Using the holomorphic anomaly equation in its polynomial form (\ref{polrec1},\ref{polrec2}) and the boundary conditions, the higher genus amplitudes can be fixed. For example at genus $2$ with the given choice of generators this gives:
\begin{eqnarray}
\mathcal{F}^{(2)}&=&\frac{1}{36000} \Big( 5 P^2 \left(37500 (S^{zz})^3+6000 (S^{zz})^2+325
   \,S^{zz}+6\right)\nonumber\\
   &&-P \left(25000 \, S^z (30
   \,S^{zz}+1)+150000 (S^{zz})^2+4400\, S^{zz}-23\right)\nonumber\\
   &&+16\,(175000 \,S+35000 \,S^z+3500\,S^{zz}-39)\Big) \, .
\end{eqnarray}
Using the polynomial generators of Ref.~\cite{Yamaguchi:2004bt}, the boundary conditions reviewed earlier as well as additional counting techniques for invariants as in Ref.~\cite{Katz:1999xq}, it was possible in Ref.~\cite{Huang:2006hq} to solve the quintic to very high genus.


\subsection*{Acknowledgments}

I would like to thank the Mathematical Sciences Center at Tsinghua University and Prof.~Shing-Tung Yau for hospitality and for the invitation to give these lectures. I would also like to thank Michael Hecht, Albrecht Klemm, Dominique L\"ange, Peter Mayr, Emanuel Scheidegger, Shing-Tung Yau and Jie Zhou for discussions and collaborations on some of the topics covered in these lectures and related subjects. I want furthermore to thank Emanuel Scheidegger and Jie Zhou for comments on the manuscript.  This work is supported by the DFG fellowship AL 1407/1-1.

\end{document}